\documentclass[twocolumn,showpacs,preprintnumbers,amsmath,amssymb,pre]{revtex4-1}
\usepackage{color}    
\usepackage{graphicx}
\usepackage{dcolumn}
\usepackage{bm}
\usepackage{subfigure}
\usepackage{amssymb}
\usepackage{multirow}
\graphicspath{{plots/}}

\renewcommand{\vec}[1]{\mathbf{#1}}

\begin{document}

% \keywords{plasma crystals, streaming effects, Yukawa balls, string formation, dynamical screening}

%\title[The ion potential in warm dense matter]{The ion potential in warm dense matter: wake effects due to streaming degenerate electrons}
%\title{Testing recently proposed ion potentials in warm dense matter against the RPA and Mermin screened potential}
\title{Statically screened ion potential and Bohm potential in a quantum plasma}

%\author{M. Bonitz, E. Pehlke, and T. Schoof}%
%\affiliation{%
%    Christian-Albrechts-Universit\"at zu Kiel, 
%    Institut f\"ur Theoretische Physik und Astrophysik, 
%    Leibnizstra\ss{}e 15, 24098 Kiel, Germany
%}

\author{Zhandos Moldabekov$^{1,2}$, Tim Schoof$^1$, Patrick Ludwig$^1$, Michael Bonitz$^1$, Tlekkabul Ramazanov$^2$}

\affiliation{
 $^1$Institut f\"ur Theoretische Physik und Astrophysik, Christian-Albrechts-Universit\"at zu Kiel,
 Leibnizstra{\ss}e 15, 24098 Kiel, Germany}

\affiliation{
$^2$Institute for Experimental and Theoretical Physics, Al-Farabi Kazakh National University, 71 Al-Farabi str.,  
 050040 Almaty, Kazakhstan
}

%\ead{ludwig@theo-physik.uni-kiel.de}

\begin{abstract}
The effective potential $\Phi$ of a classical ion in a weakly correlated quantum plasma in thermodynamic equilibrium at 
finite temperature is well described by the RPA screened Coulomb potential. Additionally, collision effects can be 
included via a relaxation time ansatz (Mermin dielectric function). These potentials are used to study the quality of 
various statically screened potentials that were recently proposed by Shukla and Eliasson (SE) [Phys. Rev. Lett. {\bf 108}, 165007 (2012)], 
Akbari--Moghanjoughi (AM) [Phys. Plasmas {\bf 22}, 022103 (2015)] and Stanton and Murillo (SM) [Phys. Rev. E {\bf 91}, 033104 (2015)] 
starting from quantum hydrodynamic theory (QHD). Our analysis reveals that the SE potential is qualitatively different from the full potential, 
whereas the SM potential (at any temperature) and the AM potential (at zero temperature) are significantly more accurate. This confirms the correctness of the recently derived [Michta {\em et al.}, Contrib. 
Plasma Phys. {\bf 55}, (2015)] pre-factor $1/9$ in front of the Bohm term of QHD for fermions. 
\end{abstract}

%52.25.Dg Plasma kinetic equations
%52.27.Aj Single-component, electron-positive-ion plasmas
%52.27.Gr Strongly-coupled plasmas
%52.65.Yy Molecular dynamics methods
%Plasma simulation, 52.65.-y
\maketitle

\section{Introduction}
Dense plasmas have recently gained growing interest due to their relevance for the interior of giant planets as well as for laser interaction with matter and inertial confinement fusion scenarios. Examples of recent experimental studies include the ultrafast thermalization of laser plasmas \cite{kluge_pop14} or free electron laser excited plasmas \cite{zastrau_prl14}, inertial confinement fusion experiments at the National Ignition Facility \cite{hurricane_nat14} and magnetized Z-pinch experiments at Sandia \cite{gomez_14}. Questions of fundamental theoretical importance are the conductivity and heat conduction, the energy loss of energetic particles (stopping power) in such a plasma, e.g. \cite{grabowski_prl13} or the temperature equilibration of the electronic and ionic components \cite{zastrau_prl14}. 

Despite recent advances in modeling and computer simulations a fully selfconsistent treatment of these, in general, highly nonequilibrium electron-ion plasmas has not been possible so far due to the requirement of the simultaneous account of electronic quantum and spin effects together with the (possibly) strong ionic correlations. The main problem here are the vastly different time scales of electrons and ions resulting from their different masses.
A possible solution of this dilemma is a multi-scale approach that has been proposed by Ludwig {\em et al.} in Ref. \cite{Patrick2}. It takes advantage of the weak electron-ion coupling that allows for a linear response treatment of the electrons. This idea has been used by Graziani {\em et al.} to decouple the electron kinetic equation using an STLS (Singwi-Tosi-Land-Sj\"olander) scheme \cite{graziani} or a recently derived extension \cite{kaehlert_pre14}.

The key of this multiscale approach is to absorb the fast electron kinetics into an effective screened potential $\Phi$ of the heavy ions with charge $Q$ where the screening is provided by the electrons via a proper dielectric function $\epsilon$, e.g. \cite{Patrick2}
\begin{equation} \label{POT_stat}
\Phi(\vec r)   = \int\!\frac{\mathrm{d}^3k}{2 \pi^2 } \frac{Q}{k^2 \epsilon(\vec k, \omega=0)} e^{i \vec k \cdot \vec r} \quad,
\end{equation}
taken in the static limit.
We note that also the effect of streaming electrons maybe important in warm dense matter. Then the frequency argument becomes $\omega = {\bf k}{\bf u}$ which leads to wake effects, that are well investigated theoretically and experimentally for dusty plasmas, e.g. \cite{Tsytovich,block12, Patrick1} and quantum plasmas, e.g. Ref.~\cite{zhandos_pre15} and references therein. However, this is beyond the scope of this paper. 

A similar but even simpler approach that is based on a quantum hydrodynamic model (QHD) and replaces the linear response potential of the ions by a simplified expression that is derived from linearized QHD (LQHD), $\Phi \longrightarrow \Phi^{\rm LQHD}$.
The first expression for $\Phi^{\rm LQHD}$ was derived by Shukla and Eliasson (SE) who predicted an attractive interaction between ions, even in the absence of streaming, $u_e = 0$ \cite{shukla_prl12}. Comparisons with density functional theory revealed that this is incorrect \cite{bonitz_pre13}. This also underlined the limitations of QHD models to weakly correlated dense plasmas \cite{bonitz_psc13,krishnaswami14}.
A recent overview and more references can be found in Ref.~\cite{Vladimirov}. 

Recently, modified expressions for $\Phi^{\rm LQHD}$ were derived by Akbari--Moghanjoughi (AM) \cite{akbari_pp15} and Stanton and Murillo (SM) \cite{stanton_pre15}. The latter is particularly interesting because it is also applicable to finite temperatures, in contrast to the SE and AM potentials that neglect thermal effects. The question arises whether the AM and SM potentials are more accurate than the SE potential and whether they allow for an extension of the applicability limits of QHD.

It is the goal of the present paper to perform such an analysis. To quantify the accuracy of these three potentials we use, as a basis, the standard static potentials of weakly correlated electrons that can be derived from quantum kinetic theory \cite{bonitz-book} by linearization in the external perturbation. This directly leads to the static potential (\ref{POT_stat}) involving the random phase approximation (RPA) for the dielectric function and the Mermin dielectric function, respectively. Both include kinetic effects, finite temperature and, in the latter case, also collisions. Being rooted in a kinetic approach, the corresponding static potentials are, by construction, more accurate than the potentials $\Phi^{\rm LQHD}$ and, thus, allow for a rigorous test of the validity of the latter.

%that are obtained from a simpler hydrodynamic approach \cite{Vladimirov}. Nevertheless, a quanitative comparison is of high fundamental and practical interest because it points out the accuracy and range of validity of the hydrodynamic potentials that are much easier to compute.

The main conclusions of our analysis are the following: 1.) the SE potential is qualitatively different from the kinetic results. 2.) 
 the SM potential exhibits very good agreement with the RPA. 3.) the AM potential is in agreement with the SM potential, at $T=0$, differing only in the notation and 4.) the SM potential is in good agreement with the RPA result even at elevated temperature.
This confirms the correctness of the recently reported \cite{michta_cpp15} correction factor $1/9$ that has to be included in front of the Bohm term of the QHD equations for fermions.

The paper is organized as follows: We first recall the RPA and Mermin dielectric functions in Sec.~\ref{s:mermin} and discuss 
their basic properties such as the reproduction of Friedel oscillations. After this we recall the SE, AM and SM potentials in Sec.~\ref{s:potential} and present numerical results that compare them to the potential $\Phi$ computed from the RPA and Mermin dielectric functions. The paper concludes with a brief discussion of the physical origin of the factor $1/9$. The main steps of its derivation are outlined in an appendix.

%----------------------
\section{Effective ion potentials screened by the RPA and Mermin dielectric functions}\label{s:mermin}
\begin{figure}[t]
\includegraphics[width=.95\linewidth]{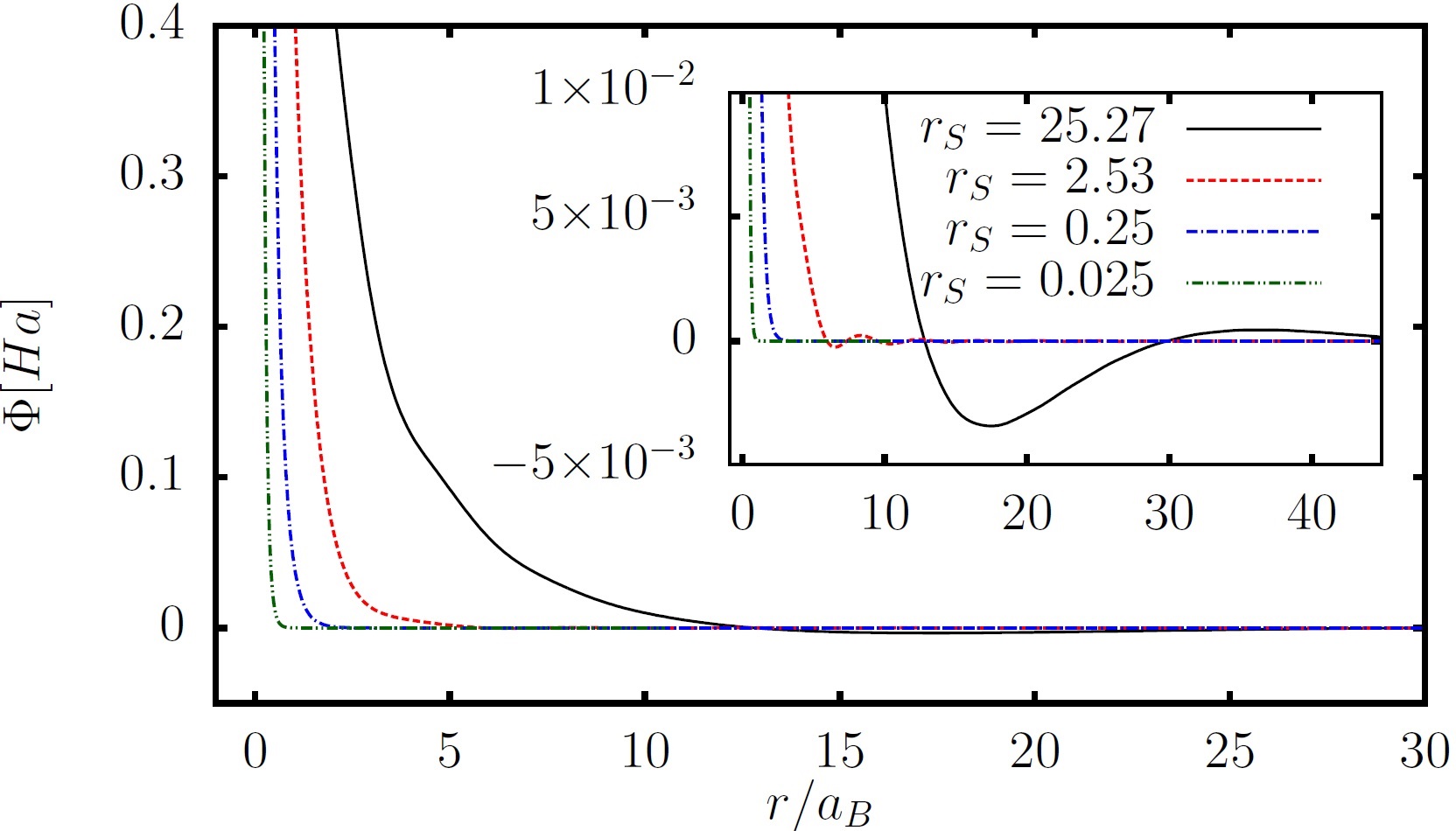}
\caption{(Color online) Effective ion potential computed from the RPA dielectric function for four densities and $T=\Theta=0$. Inset shows an 
enlargement of the potential around its zero values to resolve the Friedel oscillations.}
\label{fig:rpa_friedel}
\end{figure}
\begin{figure}[t]
\includegraphics[width=.95\linewidth]{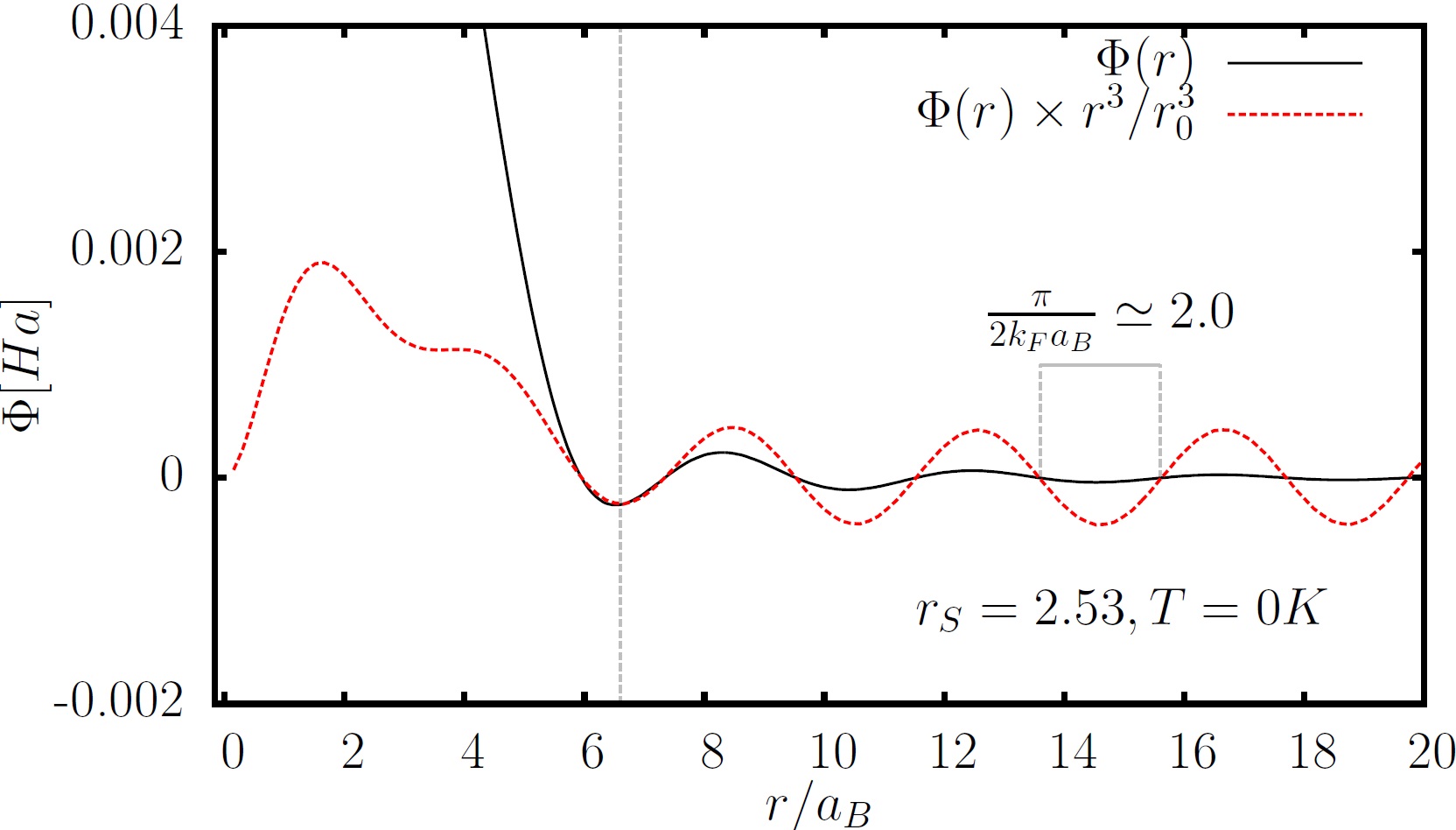}\\[2ex]
\includegraphics[width=.95\linewidth]{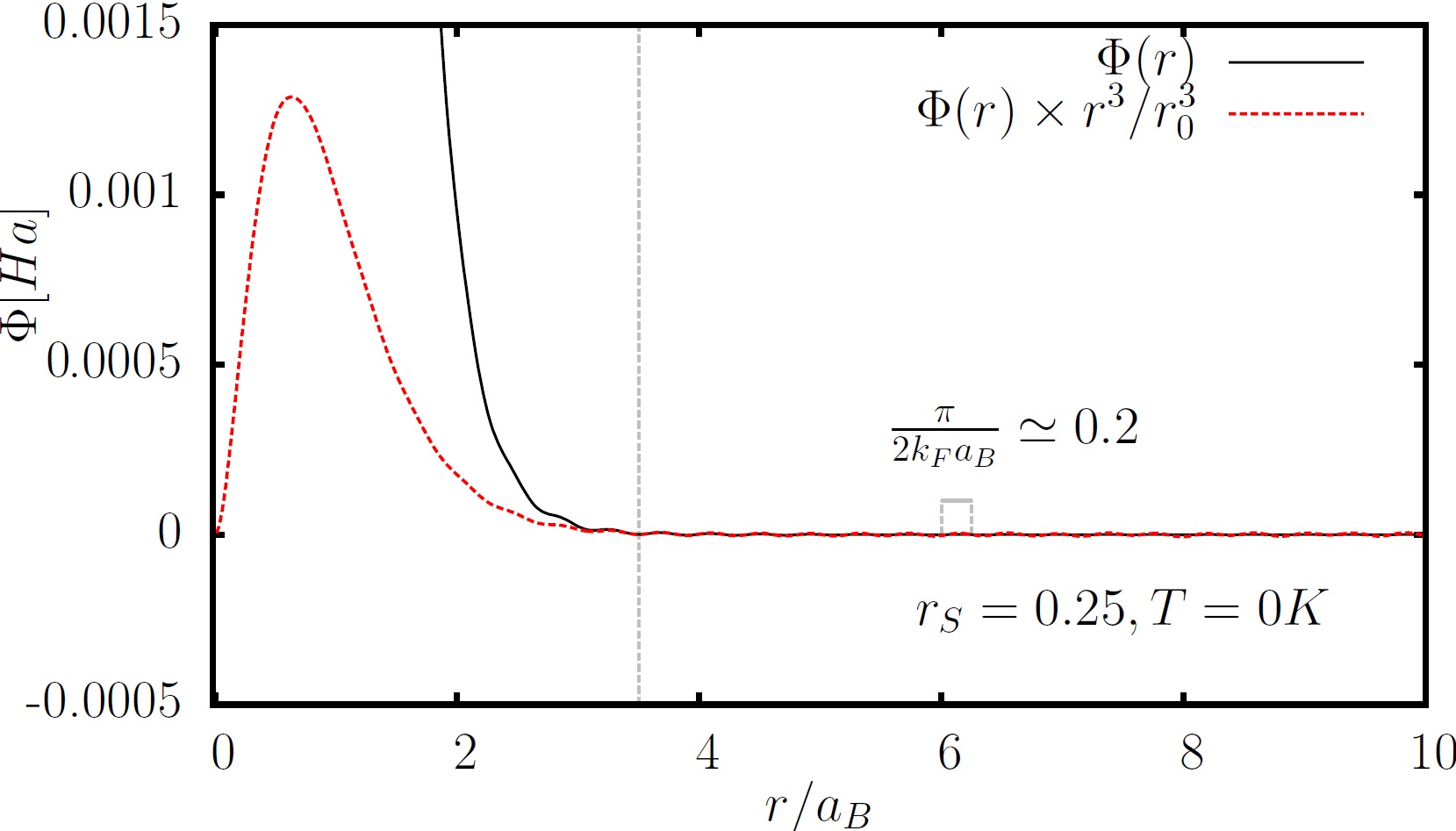}% \hfil
\caption{(Color online) Effective ion potential computed from the RPA dielectric function for two densities corresponding to $r_s=2.53$ (top) and $0.25$ (bottom) and $T=\theta=0$. 
The zoom into the small potential values confirms the correct oscillation frequency at large distances that is expected for Friedel oscillations, 
as seen from the comparison to the (red) dashed curve. Here $r_0$ corresponds to the distance indicated by a vertical dashed line. The grey box shows that the oscillation period of the full potential agrees with the known asymptotics.}
\label{fig:rpa_friedel2}
\end{figure}

The mean-field result for the electron dielectric function is given by the random phase (RPA or Hartree or quantum Vlasov)  approximation, $\epsilon_{\rm RPA}(k,\omega)$, where correlation effects are neglected, and damping of collective oscillations is entirely due to Landau damping. Temperature effects are straightforwardly included by using the corresponding Fermi function. For an overview on the final expression, see the Appendix of Ref.~\cite{zhandos_pre15}. 
The effective ion potential (\ref{POT_stat}) computed from the RPA dielectric function in the zero temperature limit is shown in Figs.~\ref{fig:rpa_friedel} and \ref{fig:rpa_friedel2} for different densities. The zero temperature RPA potential reproduces the Friedel oscillations, as is illustrated in Fig.~\ref{fig:rpa_friedel2} by comparison to the large $r$ asymptotics, $~\cos(2k_Fr)/r^3$, (red dashed line). 
%In Fig.~\ref{fig:rpa_friedel2} the distance between two zeros of the Friedel oscillations, which in agreement with a correct analytical  asymptotics , is shown.
While the oscillations are very shallow and hardly of relevance for the thermodynamics and transport properties in warm dense matter, the reproduction of the correct periodicity is a useful consistency and accuracy test of the Fourier transformation in Eq.~(\ref{POT_stat}).

For later reference we also provide the static long wavelength limit of the effective ion potential which is nothing but the Yukawa potential (\ref{POT_stat})
\begin{equation}\label{Yukawa}
\Phi_{Y}(r;n,T)=\frac{Q}{r}\, e^{-k_{Y}r},
\end{equation}
with the familiar inverse Yukawa screening length, $k_{Y}$, that interpolates between the Debye and Thomas-Fermi expressions, in the non-degenerate and zero temperature limits, respectively,
\begin{equation}\label{ksc}
k_{Y}^2(n,T)=\frac{1}{2} k_{TF}^2 \theta ^{1/2} I_{-1/2}(\beta \mu).
\end{equation}
Here $k_{TF}=\sqrt{3}\omega_{p}/v_{F}$ is the Thomas-Fermi wave number, and $I_{-1/2}$ is the Fermi integral of order $-1/2$. 
Note that the potential (\ref{Yukawa}) depends--via $k_Y$--on density and temperature.

In the warm dense matter regime we expect the electrons to be weakly to moderately coupled so that correlation effects are, in general, not negligible. 
The simplest quantum dielectric function (DF) which takes collisions into account in a conserving fashion via a relaxation time approximation
is the Mermin DF \cite{Mermin}
\begin{multline}\label{Mermin}
 \epsilon_{\rm M}(k,\omega)=1+\\ +\frac{(\omega+i\nu)[\epsilon_{\rm RPA}(k,\omega+i\nu)-1]}{\omega+i\nu[\epsilon_{\rm RPA}(k,\omega+i\nu)-1]/[\epsilon_{\rm RPA}(k,0)-1]},
\end{multline}
which involves the finite temperature RPA DF and the electron collision frequency $\nu$.
In Ref.~\cite{zhandos_pre15} various choices for the collision frequency were studied, so here we will just use a typical set of values to highlight the main trends.

%----------------------------
\section{Testing statically screened ion potentials derived from QHD} \label{s:potential}
\subsection{The Shukla-Eliasson (SE) potential}\label{ss:sep}
The Shukla-Eliasson potential has been derived from linearized QHD, below we reproduce its most recent variant \cite{shukla_prl12}.
%It contains a key parameter $\alpha^{SE}$ that is defined by $\alpha^{SE} = \hbar^2 \omega_{pe}^2 / 4m_*^2(v_*^2/3 + v_{ex}^2)^2$, 
It involves the electron plasma frequency $\omega_{pe} = (4\pi n_0 e^2 / \epsilon m_*)^{1/2}$,
[$m_*$ is the effective electron mass] and  
the electron Fermi speed $v_*=\hbar (3\pi^2)^{1/3}/m_*r_0$, where $r_0 = n_0^{-1/3}$ is proportional to the Wigner-Seitz radius.
The term $v_{ex} = (0.328 e^2 / m_* \epsilon r_0)^{1/2} \times [1+0.62 / (1 + 18.36 a_B n_0^{1/3})]^{1/2}$ can be regarded as 
(optional) correction that is supposed to account for exchange-correlation effects [it was proposed in Ref.~\cite{manfredi_08} based on the exchange-correlation potential of density functional theory]. Since the other potentials do not include such a correction and it does not qualitatively alter the potential, in the comparisons below we will use $v_{ex}=0$.

The SE potential, for $\alpha^{SE} > 0.25$, is given by \cite{shukla_prl12}:
\begin{align}
\phi^{SE>}(r;n) = \frac{Q}{r}\left[\cos(k'_- r)+ b' \sin(k'_- r )\right] e^{-k'_+ r},
\label{eq:sep}
\end{align}
with the parameter $b' = 1/\sqrt{4\alpha^{SE} - 1}$. The wave vectors are given by~\cite{kpmnote} $k'_\pm = k_s (\sqrt{4\alpha^{SE}} \pm 1)^{1/2} / \sqrt{4\alpha^{SE}}$, where $k_s = \omega_{pe} / \sqrt{v_*^2/3 + v_{ex}^2}$ is similar (but not identical) to the inverse Thomas-Fermi screening length $k_Y$, Eq.~(\ref{ksc}).
\par
The SE potential for the second parameter range, $\alpha^{SE} < 0.25$, is given by:
\begin{align}
\phi^{SE<}(r;n) = \frac{Q}{2r}\left[(1+b) \, e^{-k_+ r} + (1-b) \, e^{-k_- r}\right].
\end{align}
with the definitions $k_\pm = k_s(1\mp \sqrt{1-4\alpha^{SE}})^{1/2} / \sqrt{2 \alpha^{SE}}$ and $b = 1/\sqrt{1 - 4\alpha^{SE}}$.

Note that the Shukla-Eliasson potential is a zero-temperature approximation, and the potential depends only on density. Numerical results are included in Fig.~\ref{fig:all_vs_rpa0}.
%------------
%
\subsection{The Akbari-Moghanjoughi (AM) potential}\label{ss:amp}
Here we reproduce the electrostatic potential derived by Akbari-Moghanjoughi (we use the most recent corrected version, Ref.~\cite{akbari_pp15}).
This potential has the same mathematical form as the SE potential $\phi^{SE>}$, for $\alpha^{AM} > 0.25$:
\begin{align}
\phi^{AM>}(r;n) = \frac{Q}{r}\left[\cos(k'_- r)+ b' \sin(k'_- r )\right] \, e^{-k'_+ r}.
\label{eq:amp>}
\end{align}
The difference lies in the definition of the parameter $\alpha^{AM} = \hbar^2 \omega_{pe}^2 / 36m_*^2(v_*^2/3)^2$  and of the wave numbers~\cite{kpmnote} $k'_\pm = k_{TF} (\sqrt{4\alpha^{AM}} \pm 1)^{1/2} / \sqrt{4\alpha^{AM}}$ where $k_{TF} = \omega_{pe} / \sqrt{v_*^2/3}$ is the inverse Thomas-Fermi screening length. These formulas are connected to the SE versions by $v_{ex} = 0$ and an additional factor of $1/9$ in the definition of $\alpha^{SE}$. 

The potential for $\alpha^{AM} < 0.25$ is not given in Ref.~\cite{akbari_pp15}. Assuming the same relation to the SE result it has the form
\begin{align}
\phi^{AM<}(r;n) = \frac{Q}{2r}\left[(1+b) \, e^{-k_+ r} + (1-b) \, e^{-k_- r}\right],
\label{eq:amp<}
\end{align}
with $b = 1/\sqrt{1 - 4\alpha^{AM}}$ and $k_\pm = k_{TF}(1\mp \sqrt{1-4\alpha^{AM}})^{1/2} / \sqrt{2 \alpha^{AM}}$.

As the SE potential, this is a zero-temperature result which, therefore, depends only on the density. Numerical results are included in Fig.~\ref{fig:all_vs_rpa0}.

%-----------------------
\subsection{The Stanton-Murillo (SM) potential}\label{ss:smp}
Here we reproduce the screened potential of Stanton and Murillo following the latest corrected version \cite{stanton_pre15}.
Again, there are two cases. For $\alpha^{SM}>1$ the potential is given by \cite{SM_formulas_note}
\begin{align}\label{eq:sm>}
\phi^{SM>}(r;n,T) = \frac{Q}{r}\left[\cos(k'_- r)+ b' \sin(k'_- r )\right] \, e^{-k'_+ r},
\end{align}
with $\alpha^{SM} = 3 \sqrt{8 \beta} \lambda {I'}\!\!_{-1/2}(\eta_0) / \pi$, $\lambda = 1/9$, $b' = 1/\sqrt{\alpha^{SM} - 1}$, and $k'_\pm = k_{TF}(\sqrt{\alpha^{SM}} \pm 1)^{1/2} / \sqrt{\alpha^{SM}}$. $I_p(\eta) = \int_0^\infty dx\; x^p / (1 + e^{x-\eta})$ denotes the Fermi integral and $I'_p(\eta)$ its derivative with respect to $\eta$. $\eta_0$ is determined by the normalization, $n_0 = \sqrt{2} I_{1/2}(\eta_0) / \pi^2\beta^{3/2}$ with the inverse temperature $\beta$. The inverse Thomas-Fermi screening length for finite temperatures is given as $k_{TF} = (4I_{-1/2}(\eta_0)/\pi\sqrt{2\beta})^{1/2}$.

A second version of the potential follows for the case $\alpha^{SM} < 1$:
\begin{align}
\phi^{SM<}(r;n,T) = \frac{Q}{2r}\left[(1+b)\, e^{-k_+ r} + (1-b)\, e^{-k_- r}\right],
\end{align}
where $b = 1/\sqrt{1 - \alpha^{SM}}$ and $k_\pm = k_{TF}(1\mp \sqrt{1-\alpha^{SM}})^{1/2} / \sqrt{\alpha^{SM}/2}$. 

\begin{figure}[h]
%\hspace{2mm}
%\begin{minipage}{0.48\linewidth}
%\hspace{-2mm}
\includegraphics[width=.88\linewidth]{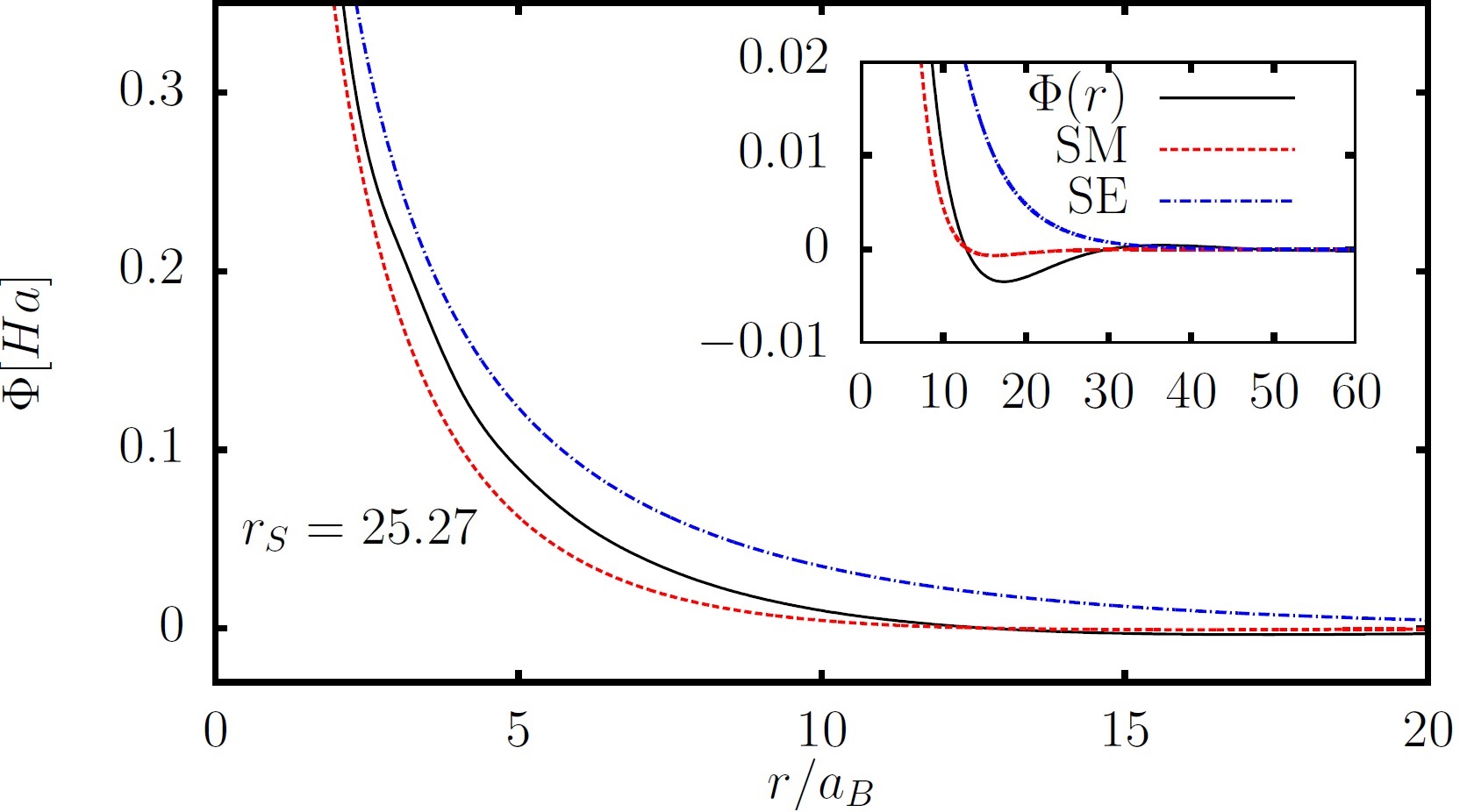}\\[2ex]
\includegraphics[width=.88\linewidth]{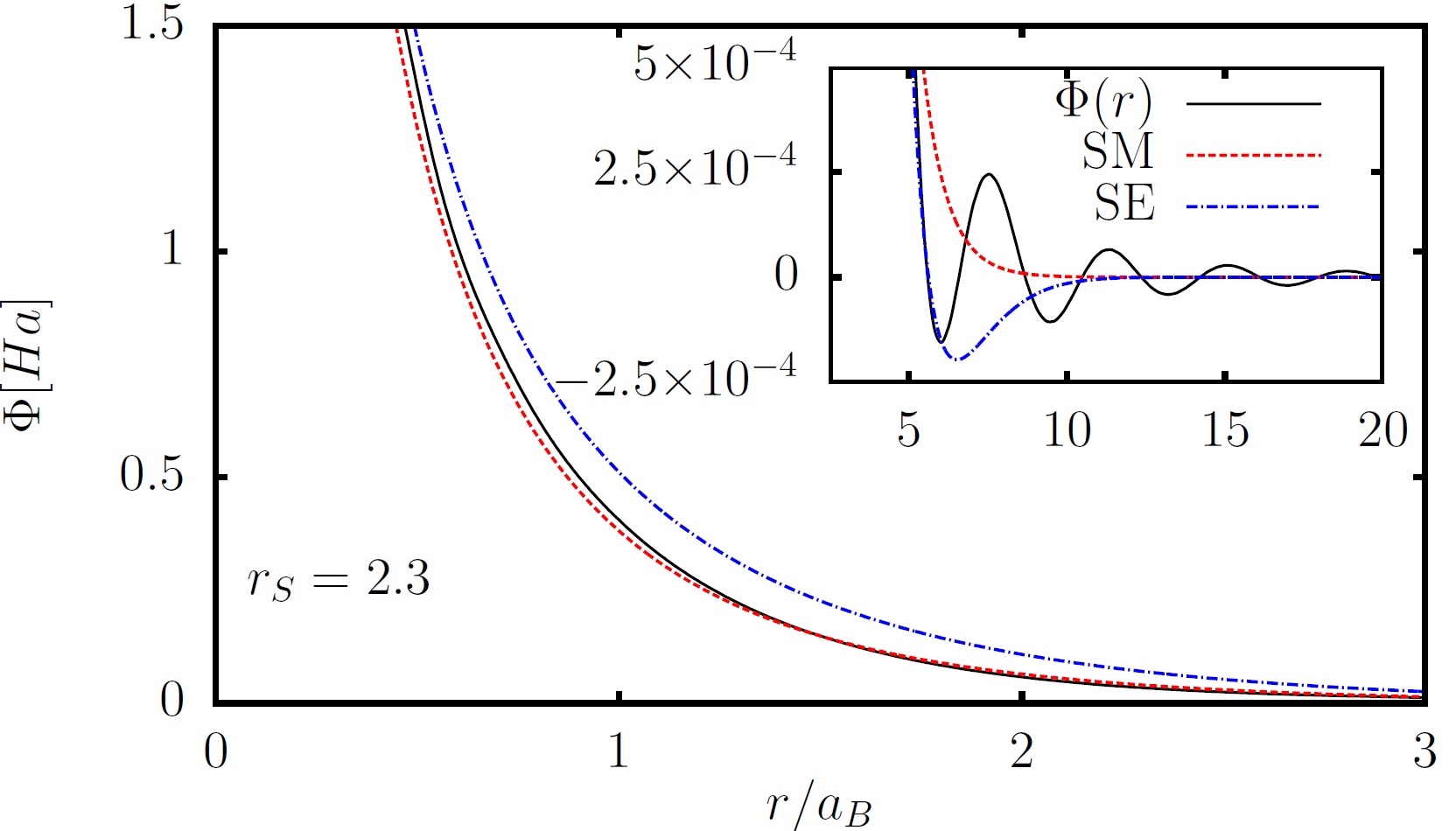}\\[2ex]
\includegraphics[width=.88\linewidth]{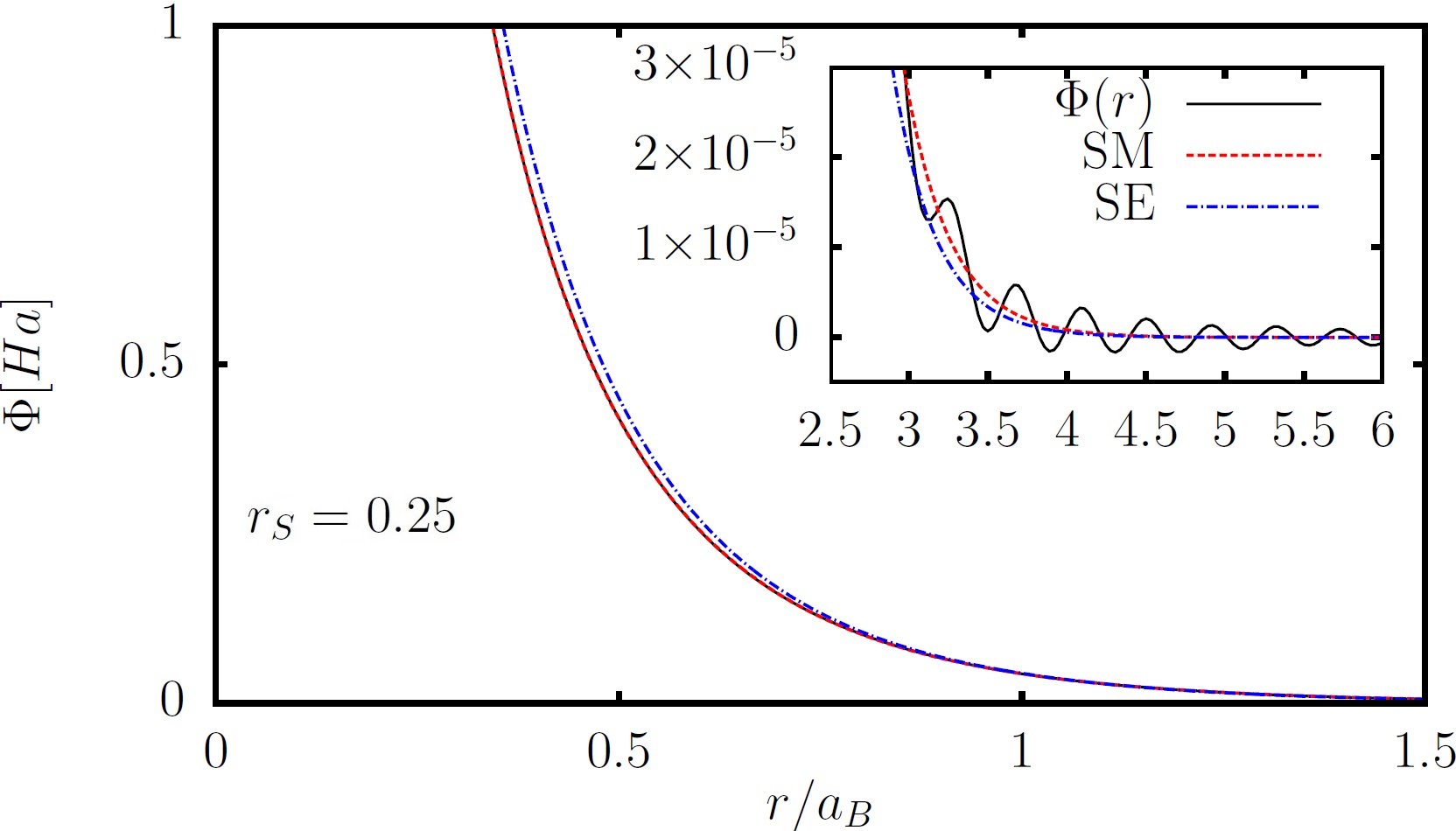}
\caption{(Color online) Comparison of the Shukla-Eliasson (SE, blue dashes) and Stanton-Murillo (SM, red dots) [it coincides to AM for $T=0$] potentials to the zero-temperature full RPA potential (full black line) for three densities. Top: $r_s=25.27$, center: $r_s=2.3$, bottom: $r_s=0.25$. The SM potential shows increasing deviations from the RPA potential $\Phi$ for decreasing density, cf. top and middle figures.
At high density (bottom figure) the SM potential exhibits very good agreement with the RPA potential. 
The SE potential exhibits large deviations from the RPA at all densities. 
Insets show the behavior for large distances and confirm that all models (except for the RPA) are unable to describe the Friedel oscillations correctly.}
\label{fig:all_vs_rpa0}
\end{figure}

We note that the SM expressions are identical to the AM version in the limit $T=0$, if $\alpha^{AM}$ is rescaled  by a factor of $4$, 
$\alpha^{SM} = 4 \alpha^{SM}(T=0)$. Numerical results are included in Figs.~\ref{fig:all_vs_rpa0} and \ref{fig:ms_rpa_t}.

\subsection{Numerical test of the potentials $\Phi^{\rm LQHD}$ from linearized QHD}\label{ss:comparison}
We now present a numerical comparison of the three LQHD potentials to the RPA potential (\ref{POT_stat}). In the following, all results are given in Hartree atomic units. Since the AM potential conincides with the SM potential taken at $T=0$ we will, in the following, not distinguish both potentials. We use the shortcut ``SM'' for both cases. 

\subsubsection{Zero temperature}
To begin the analysis, we recall that the SE, SM [and AM] ion potentials are all derived by applying the long-wavelength limit to the full potential. 
Correspondingly, in Fourier space one should expect accurate results only for small wave numbers,  $k \ll 2k_{F}$, where $k_{F}=(3\pi^2 n)$ is the Fermi wavenumber.

For a comprehensive comparison we consider a broad density range, covering 6 orders of magnitude with the three values of the Brueckner parameter,
 $r_{S}=25.27, 2.3, 0.25$ (similar values were studied in Ref.~\cite{akbari_pp15}). The first observation, cf. Fig.~\ref{fig:all_vs_rpa0}, is that all four potentials show the same overall behavior with the deviations growing when the density is reduced. The next observation is that the SM potential conincides with the RPA potential at large distances: for $r/a_{B} \gtrsim 12, 1.2, 0.9$, for the $r_s=25.57, 2.3$ and $0.25$, respectively. However, a closer inspection of the large distance behavior shows that substantial deviations remain: none of the three approximate potentials reproduces the Friedel oscillations of the RPA potential, see also Figs.~\ref{fig:rpa_friedel} and \ref{fig:rpa_friedel2}. Nevertheless, the SM potential provides a good fit to the RPA potential if one discards the (small) Fourier component with $k=2k_F$.

%correspondingly the SM and AM potentials, which take into account
%the collective effects due to electrons in  the long-wavelength limit, describe correctly the charge screening at $r/a_{B}>7, 0.6, 0.06$.
%It is demonstrated clearly in Fig.~\ref{fig:all_vs_rpa0} (top and middle), where at zero temperature limit the SM potential and the AM potential coincide exactly. At higher density it is less illustrative  due to the Coulomb singularity.
%However, the resolution limit of the SM and AM potentials at high densities can be demonstrated via the manifestation of the Friedel oscillations which appear
%at wavenumber $k=2k_{F}$ as it is seen from their asymptotics $~\cos(2k_{F}r)$. This is illustrated in Fig.~\ref{fig:all_vs_rpa0} via the zoom into small potential values.
Consider now the case $r_s=2.3$, cf. middle part of Fig.~\ref{fig:all_vs_rpa0}. Here one clearly recognizes the difference between the SM (and AM) potentials, compared to the SE potential. While the SE potential exhibits strong deviations from the RPA potential, the SM potential is very close to the RPA result. At the highest density, $r_s=0.25$, bottom figure, the deviations between SM and SE are very small. But this is not surprising as, in this case, the Bohm term in the QHD equations is negligible, so the influence of the different prefactors of this term ($1$, for SE, versus $1/9$, for SM) is insignificant. Finally, at the lowest density, $r_s=25.27$, top figure, all three potentials show substantial deviations from the RPA for small distances below $10 a_B$, although the deviations of the SE potential are significantly larger, in particular for larger distances. However, at these low densities, the plasma is strongly correlated and the RPA potential itself is not applicable.

%On the other hand, the SM and AM potentials, within their applicability range, correctly describe charge screening due to degenerate electrons in contrast to 
%the SE potential which fails to reproduce a correct screening effect at all values of the considered density. This fact indicates the correctness of the SM and AM potentials, which in the case of the SM potential includes 
%the factor $1/9$ in the gradient correction to the kinetic energy and in the case of the AM potential takes into account the factor $1/9$ by correctly
%expanding the inverse of the RPA dielectric susceptibility, whereas the SE potential was derived neglecting this factor. 
Thus, from the zero-temperature behavior we conclude that the SM (and AM) potential shows a substantial improvement over the SE potential, confirming the correctnesss and importance of the prefactor $1/9$ in front of the Bohm term \cite{michta_cpp15}. In fact, we will show in Sec.~\ref{s:dis} that this coefficient is not a free parameter but, for the description of the long-wavelength properties of the plasma (such as the statically screened potential) the value $1/9$ follows rigorously.

\subsubsection{Finite temperature}
Quantum hydrodynamics, as used in the plasma physics community, is a zero-temperature theory, so the results of SE \cite{shukla_prl12} for the effective ion potential and the recent correction by Akbari-Moghanjoughi \cite{akbari_pp15} are restricted to the ground state. At the same time, for relevant applications to dense plasmas, warm dense matter or laser plasmas, finite temperature effects are usually non-negligible. It was shown in Refs.~\cite{michta_cpp15, stanton_pre15} that the QHD equations are directly linked to the Thomas-Fermi theory which allows for a straightforward incorporation of finite-temperature effects. This link was exploited in Ref.~\cite{stanton_pre15} to derive a statically screened potential for arbitrary temperatures. 

\begin{figure}[h]
%\begin{minipage}{0.48\linewidth}
%\hspace{-2mm}
\includegraphics[width=.95\linewidth]{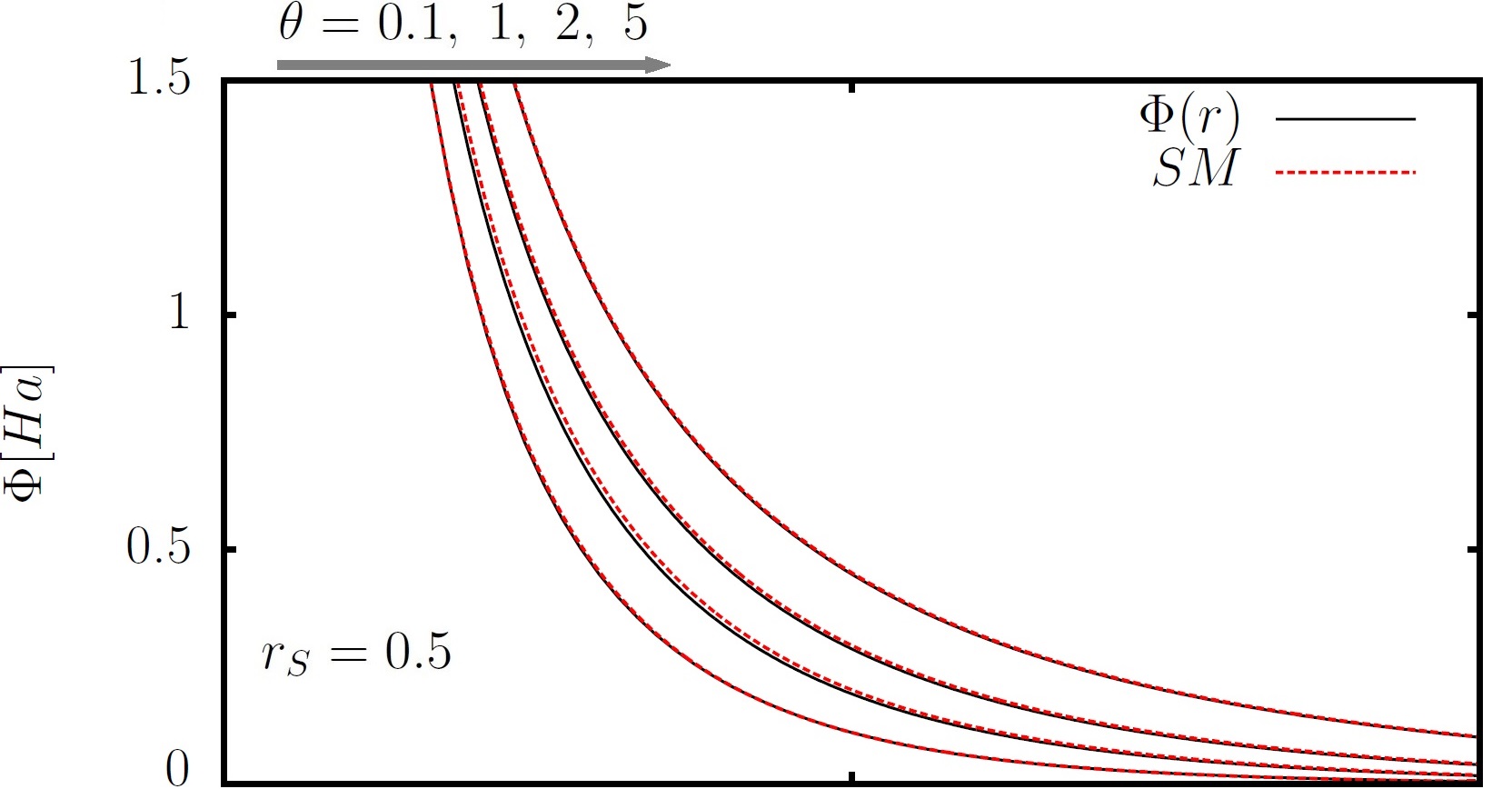}
\includegraphics[width=.95\linewidth]{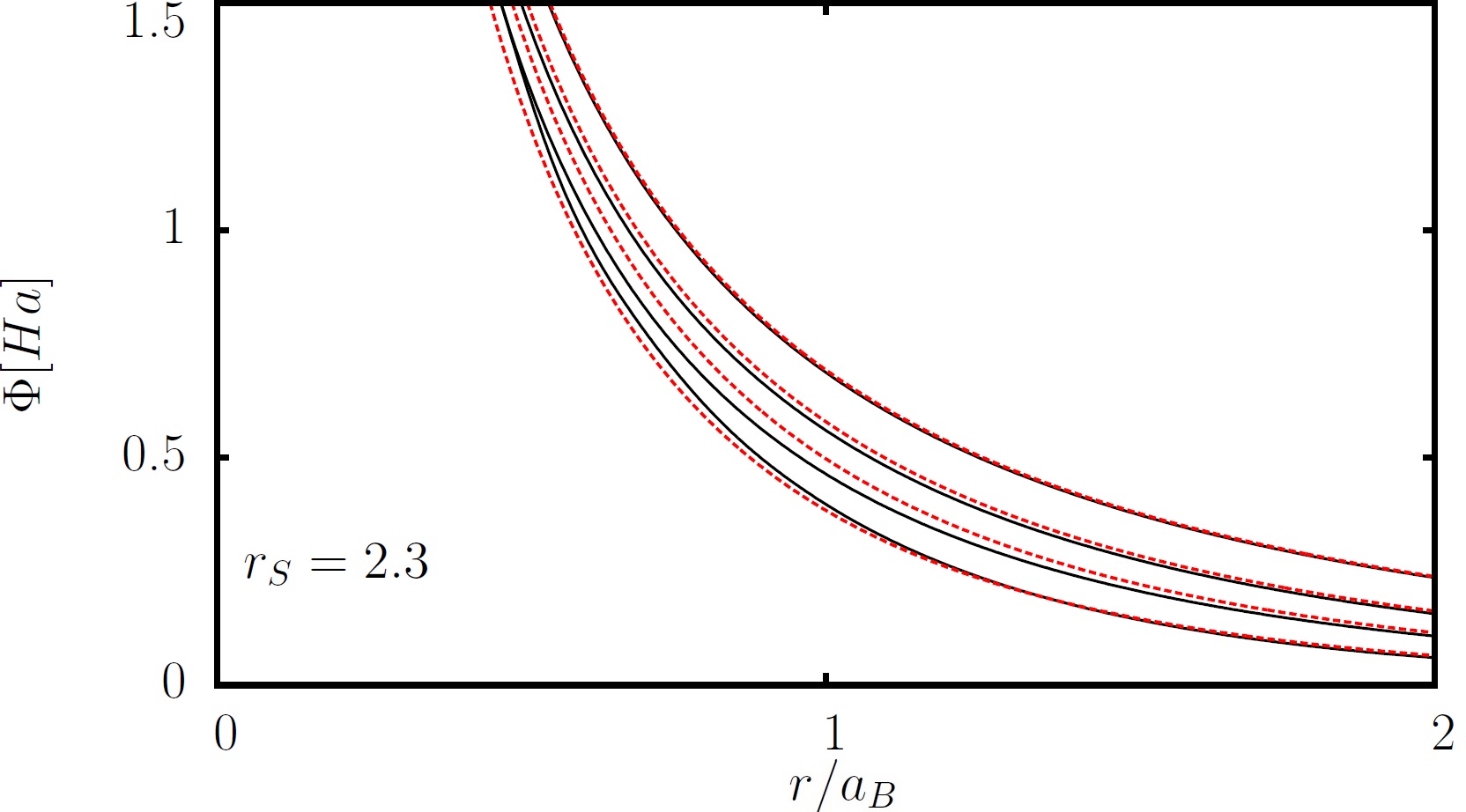}
\caption{(Color online) Comparison of the Stanton-Murillo potential to the RPA potential for various temperatures at a fixed density. Top: $r_{s}=0.5$, bottom: $r_s=2.3$. As it is seen the SM potential accurately describes the screening effect at high temperatures and densities.}
\label{fig:ms_rpa_t}
\end{figure}

It is now interesting to compare the corresponding 
finite-temperature SM potential to the RPA potential at the same temperature. This is done in Fig. ~\ref{fig:ms_rpa_t} for different
values of the degeneracy parameter $\theta=k_{B}T/E_{F}$ at constant density. The figure shows that, for high density, $r_s=0.5$, the agreement is excellent, for all temperatures. This is not surprising since, at these densities gradient corrections (the Bohm term of QHD) are small, as noted above. Here the temperature dependence is governed by the one of the ideal Fermi gas (Fermi pressure). More interesting is the comparison at lower density, $r_s=2.3$, bottom part of Fig.~\ref{fig:ms_rpa_t} where correlation effects (and the Bohm contribution) are significant. Here the agreement between SM and RPA potentials is slightly worse and improves with increasing temperature. This is due to smoothening of the Friedel oscillations with increased temperature.

Overall, we conclude that the agreement between the SM potential and the RPA, even at finite temperature, is very impressive. This confirms again that the SM potential correctly captures the long-wavelength properties of the static RPA potential (\ref{POT_stat}).

\begin{figure}[t]
%\begin{minipage}{0.48\linewidth}
%\hspace{-2mm}
\includegraphics[width=.95\linewidth]{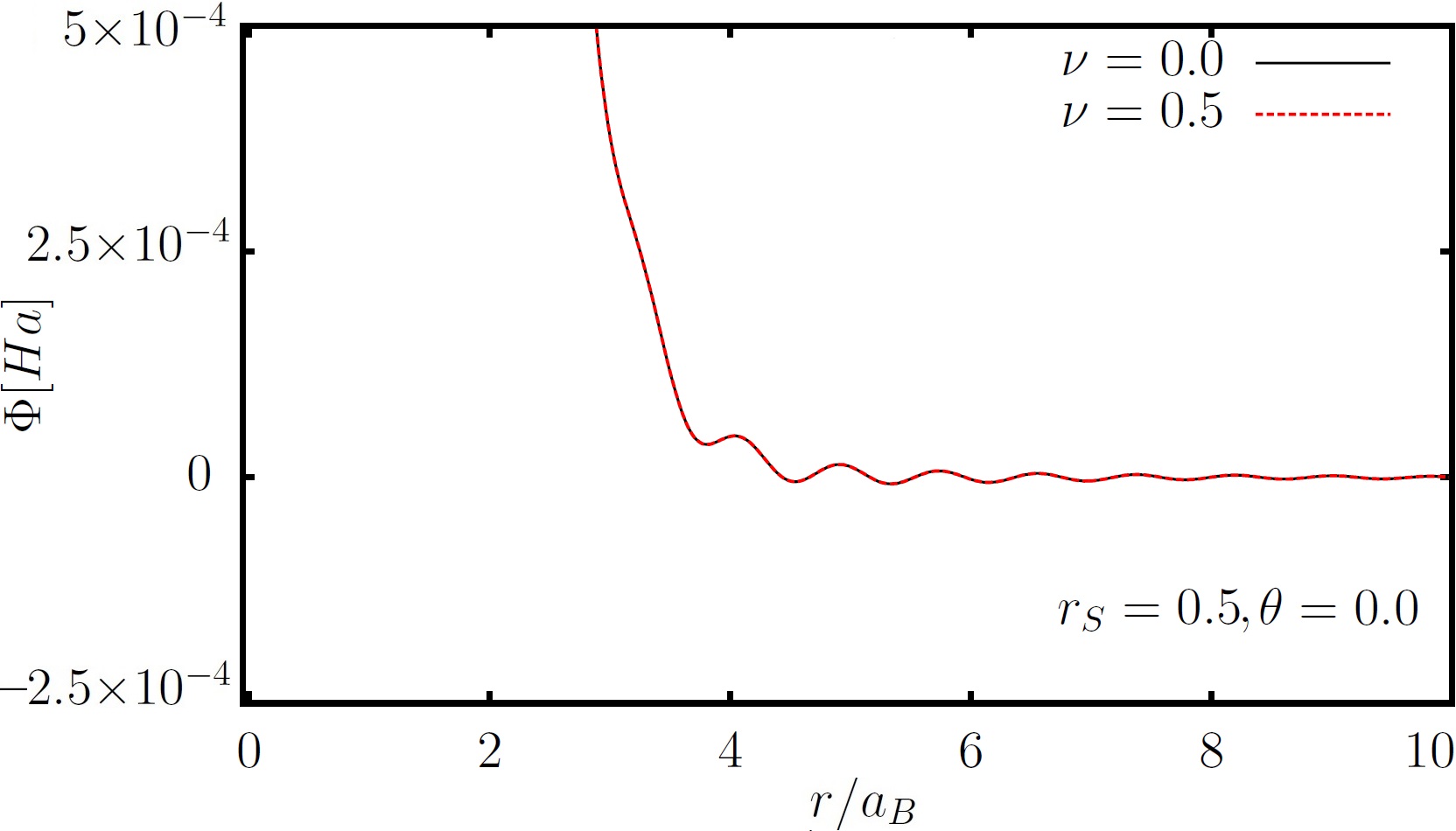}
\caption{(Color online) Ion potential at a fixed density, $r_s=0.5$ and two collision frequencies (in units of the plasma frequency). The behavior is representative for other densities and temperatures and indicates that the electron collisions do not affect the static screened potential, within the relaxation time approximation (Mermin model).}
\label{fig:ms_mermin_t}
\end{figure}
Finally, let us analyze the effect of collisions (correlations) on the dielectric function. It is well known that these effects become increasingly 
important outside the weak-coupling limit, i.e. for $r_s \gtrsim 1$. Yet it is not clear, a priori, how important these effects are for the static ion potential. We, therefore, computed the static potential (\ref{POT_stat}) with the Mermin dielectric function for a broad range of temperatures, densities and collision frequencies. A typical result is shown in Fig.~\ref{fig:ms_mermin_t} where the RPA potential is compared to the Mermin dielectric function with the  collision frequency $\nu = 0.5 \omega_{pl}$, at zero temperature. The differences are extremely small, and the same behavior is observed at other densities and elevated temperatures. The explanation is, of course, that correlation effects are particularly relevant at short distances where the Coulomb potential is large anyway. We note that the situation is very different in streaming plasmas that are out of equilibrium. Here scattering effects have an important impact on the effective ion potential \cite{ludwig_cpp14}.

\section{Summary and Discussion} \label{s:dis}
\subsection{Accuracy of the LQHD screened potentials}
We have presented an analysis of three recently proposed static ion potentials in quantum plasmas and compared them to the static limit of the RPA potential (\ref{POT_stat}). The first of these potentials (SE) was the one derived by Shukla and Eliasson \cite{shukla_prl12} who predicted an attractive minimum of the potential [cf. middle part of Fig.~\ref{fig:all_vs_rpa0}] that would be responsible for ordering of ions in dense plasmas. Our analysis reveals that no such minimum exists in the RPA potential, so this minimum should be regarded an artefact of the used LQHD model \cite{bonitz_pre13}.

The comparison with the two other potentials, the ones of Stanton and Murillo (SM) \cite{stanton_pre15} and of Akbari-Moghanjoughi  (AM) \cite{akbari_pp15}
which do not exhibit a comparable minimum reveals the origin of this discrepancy (see below). The SM (and AM) potential exhibits very good agreement with the RPA potential at $T=0$ indicating that it correctly captures the long-wavelength properties of the RPA, in contrast to the SE potential. 

\subsection{Implications for the Bohm potential for quantum plasmas}
Since the only difference between the SE potential (\ref{eq:sep}) and the SM (and AM) potential (\ref{eq:sm>}) is the different prefactor of the Bohm term in the underlying QHD equations, the good accuracy of the latter indicates that the choice of the prefactor $1/9$ is the correct one. In fact, this question is easily answered by recalling the one-to-one correspondence between QHD and Thomas-Fermi theory with gradient corrections \cite{stanton_pre15,michta_cpp15}. In Thomas-Fermi theory and, more generally, in density functional theory, it is well known that an expansion of the (non-interacting) kinetic energy (together with exchange contributions) in a power series in terms of density gradients converges to the full Hartree-Fock energy \cite{engel-dreizler89}. In the present case, we start from the statically screened potential (\ref{POT_stat}) in RPA, so we should expect agreement from a gradient expansion of the Thomas-Fermi kinetic energy $T[n]$ without exchange and correlation corrections. Since the Bohm term of QHD corresponds just to the first non-vanishing gradient correction, $T_2$, (quadratic in density gradients, see Appendix), the pre-factors of both are linked inseparably via the equation  \cite{michta_cpp15}
\begin{equation}
  \frac{\delta T_{2}[n;\gamma]}{\delta n} = \gamma\, \frac{\hbar^2}{8m}\left( \left| \frac{\nabla n}{n} \right|^2 - 2 \frac{\nabla^2 n}{n} \right) 
 = \gamma \,V^{\rm MH}_B[n],
\label{eq:grad=vb1}
\end{equation}
where $V^{\rm MH}_B$ is the Bohm potential for quantum systems with Bose statistics. For the case of quantum plasmas (fermions) the same potential (i.e. $\gamma=1$) was postulated by Manfredi and Haas \cite{manfredi-haas01}. However, the correct value (at $T=0$) is $\gamma=1/9$ which is clearly confirmed by the comparison of the SE and SM potentials to the RPA in the present work.
 
%What remains is to establish the
%relation between the coefficient $\gamma$ and the coeffient $a_2$ of the expansion (\ref{term_2l}). In the following we prove that, for zero temperature, to lowest order in the correlation energy, $a_2 = \frac{\hbar^2}{72 m}$ which means that $\gamma = 1/9$.

\subsection{Derivation of the first gradient correction to the kinetic energy for fermions. The coefficient $\gamma=1/9$}
Let us briefly discuss the origin and validity of the prefactor $1/9$. 
This, coefficient  was first obtained by  Kompaneets \cite{Kompaneets} and Kirzhnitz \cite{Kirzhnitz}, almost 60 years ago, who used a Matsubara Green functions method to obtain the gradient correction to the Thomas-Fermi energy functional. In the same year Golden \cite{Golden} 
derived this factor by expanding the density matrix of a many-electron system. Later Hohenberg and Kohn \cite{Kohn} 
presented a systematic derivation in terms of a gradient expansion to all orders, on the basis of the electron polarization function for zero temperature.
Mermin \cite{Mermin2} extended the Hohenberg and Kohn method to finite temperature
but did not give a derivation of the gradient correction at finite temperature. This was done by Krizhnitz {\em et al.} \cite{kirzhnits75} and by Perrot \cite{Perrot} who minimized the grand canonical potential using the known form of the free energy of the electrons. At last, the Feynman path-integral
method was used by Yang \cite{Yang} to obtain the single-particle Green function and to prove the factor $1/9$.

Over the recent decades occasionally additional derivations have appeared. In particular, the recent analysis of Akbari-Moghanjoughi \cite{akbari_pp15} essentially reproduces earlier derivations based on the calculation of the long-wavelength limit of the inverse of the Lindhard dielectric function, $\epsilon^{-1}_{\rm RPA}$ at $T=0$, whereas the analysis of Stanton and Murillo \cite{stanton_pre15} directly starts from Perrot's result \cite{Perrot}.

We also note that there exists a different approach to the gradient corrections in Thomas-Fermi theory where the prefactor of the respective term is used as a fit parameter to reproduce improved simulations (that go beyond the RPA), e.g. \cite{xu_hansen}. Similarly, for the case of electrons bound in atoms, the prefactor can be optimized to better match the Hartree-Fock ground state energy, yielding $\gamma=0.2$ \cite{Yonei65, Gross_Dreizler79}. However, in the context of the screened potential of free charged particles in a plasma one has to use the static long-wavelength limit of the inverse dielectric function which is exactly given by the RPA, so there exists no freedom of choice, see Sec.~\ref{ss:p-expansion}.

%For the benefit of readers we give, in the Appendix, a systematic derivation of the gradient correction to the Thomas Fermi energy. There we also include new results for the higher order gradient corrections at finite temperature.
%
\subsection{Wavenumber expansion of the inverse polarization function and inverse dielectric function. Improved static ion potential}\label{ss:p-expansion}
A key input for the gradient correction of the kinetic energy is the wavenumber expansion of the inverse of the static density response function, i.e. of the longitudinal polarization function $\Pi({\bf k}, \omega=0)$. The result is (for the derivation, see the Appendix)
\begin{align}
 {\tilde K}({\bf k}) = \frac{1}{2\Pi({\bf k})} = {\tilde a}_0
%+{\tilde a}_1 \vec{k}
+{\tilde a}_2 \vec{k}^2 + {\tilde a}_4 \vec{k}^4+....
\label{eq:pinv_exp}
\end{align}
This is a completely general result where the whole microphysics is contained in the value of the expansion coefficients. From this we can directly obtain systematic approximations for the statically screened ion potential (\ref{POT_stat}) without recourse to the connection between Thomas-Fermi theory and QHD, Eq.~(\ref{eq:grad=vb1}). Indeed, what is needed to evaluate (\ref{POT_stat}) is the inverse of the dielectric function which is related to the polarization function by \cite{bonitz-book}
\begin{equation}
\epsilon({\bf k},0) = 1 - \frac{4\pi e^2}{k^2}\Pi({\bf k}).
\label{eq:df}
\end{equation}
From this and Eq.~(\ref{eq:pinv_exp}) we obtain the desired wavenumber expansion of the inverse static dielectric function [and of the screened potential (\ref{POT_stat})]
\begin{align}
\label{eq:eps_inv_exp}
 \epsilon^{-1}({\bf k},0) = \frac{2k^2 \left({\tilde a}_0  + {\tilde a}_2 k^2  + {\tilde a}_4 k^4 + \dots \right)}{2k^2 \left({\tilde a}_0  + {\tilde a}_2 k^2  + {\tilde a}_4 k^4 + \dots \right) - 4\pi e^2}
\end{align}

Using, as the lowest order approximation, the RPA polarization, we obtain for the first five non-vanishing coeffcients  (see Appendix)
\begin{align}
 {\tilde a}_0(n,T) &= - \frac{2\pi e^2}{\kappa_Y^2(n,T)},
\\
 \frac{{\tilde a}_2(n,T)}{{\tilde a}_0(n,T)} &= -\frac{b_1}{4 k_F^2},
 \\
 \frac{{\tilde a}_4(n,T)}{{\tilde a}_0(n,T)} &=\frac{b_1^2-b_2}{16 k_F^4},
\label{eq:a4}
 \\
 \frac{{\tilde a}_6(n,T)}{{\tilde a}_0(n,T)} &=\frac{-b_1^3+2b_1b_2-b_3}{64 k_F^6},
 \\
 \frac{{\tilde a}_8(n,T)}{{\tilde a}_0(n,T)} &= \frac{b_1^4-3b_1^2b_2+b_2^2+2b_1b_3-b_4}{256 k_F^8},
\label{eq:a8}
\end{align}
where the $b_i$ involve Fermi integrals of different orders [see Eq.~(\ref{eq:bi}) of the Appendix)]. Note that the finite temperature 
results for ${\tilde a}_4, {\tilde a}_6$ and $ {\tilde a}_8$ are given here for the first time.
%$\eta=\mu/k_BT$, and
%\begin{equation}
%\label{b}
% b_i\left(n,T\right)=\frac{\Theta^{-i}}{2i+1} \frac{I_{-i-1/2}(\eta)}{I_{-1/2}(\eta)}.
%\nonumber
%\end{equation}
This result allows one to systematically derive approximations for the screened potential by considering the long-wavelength limit, $k \to 0$ of the inverse dielectric function (\ref{eq:eps_inv_exp}). 

{\bf Lowest order result:} The lowest order is given by neglecting all $k$-dependent terms in the parantheses of Eq.~(\ref{eq:eps_inv_exp}), ${\tilde a}_2={\tilde a}_4=\dots 0$, and we obtain
\begin{align}
 \epsilon^{-1}_0({\bf k},0) = \frac{k^2}{k^2 + \kappa_Y^2}.
\label{eq:eps_0}
\end{align}
Inserting this into Eq.~(\ref{POT_stat}), immediately yields the Yukawa potential (\ref{Yukawa}). 

\begin{figure}[t]
\includegraphics[width=.85\linewidth]{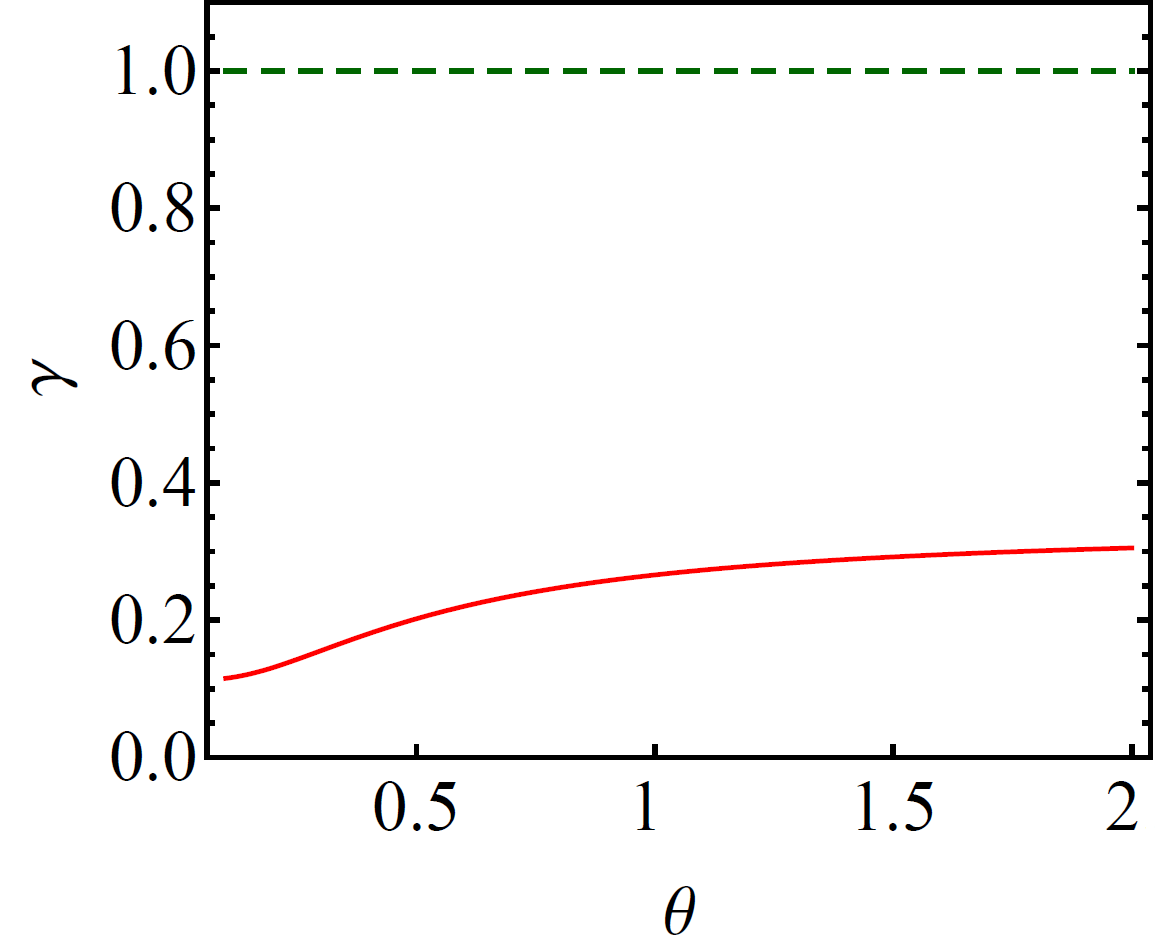}
\caption{(Color online) prefactor $\gamma$ in front of the Bohm potential $V_B^{\rm MH}$, cf. Eq.~(\ref{eq:grad=vb1}), for fermions as a function of temperature. 
In the zero-temperature limit, $\gamma$ approaches $1/9$, and it increases monotonically to  $1/3$ \cite{Perrot}. In contrast, for bosons or spinless particles this factor equals $1$ (dashed line), as in the case of a single quantum particle \cite{michta_cpp15}.}
\label{fig:alpha_3}
\end{figure}

{\bf Second order result:} The next order is obtained by retaining the terms quadratic in $k$,
\begin{align}
 \epsilon^{-1}_2({\bf k},0) = \frac{k^2\left( 1 + \frac{ {\tilde a}_2 }{ {\tilde a}_0 } k^2 \right)}{k^2 + \kappa_Y^2 + \frac{ {\tilde a}_2 }{ {\tilde a}_0 } k^4}.
\label{eq:eps_2}
\end{align}
The result for ${\tilde a}_2$ is (see Appendix)
\begin{align}\label{eq:a2-tilded}
{\tilde a}_2\left(n,T\right) &= \frac{\hbar^2 I_{-3/2}(\eta)}{36m_e n \Theta^{3/2} I_{1/2}^2(\eta)}.
\end{align}
In the ground state, $T=0$, this coefficient becomes ${\tilde a}^{(0)}_2(n,T) =  - \frac{1}{9} \frac{\hbar^2}{8m}$.
Equation (\ref{eq:eps_2}) is the approximation used by Stanton and Murillo \cite{stanton_pre15} and (for $T=0$) by
Akbari--Moghanjoughi  (AM) \cite{akbari_pp15}. The approximation of Shukla and Eliasson is obtained for $T=0$
by using ${\tilde a}_2^{(0)}$ without the prefactor $\gamma(0)=1/9$. This coefficient yields the ratio of the corresponding results for fermions and bosons (or spinless particles) \cite{michta_cpp15}.
From the temperature dependence of ${\tilde a}_2$ we can extract the temperature dependence of the coefficient $\gamma$ which is displayed in Fig.~\ref{fig:alpha_3}. $\gamma$ increases monotonically from $1/9$ to $1/3$ \cite{Perrot, kirzhnits75}, for temperatures large compared to $E_F$. 
Evidently this factor is crucial for the correct treatment of fermions within Thomas-Fermi theory or in quantum hydrodynamics.

\begin{figure}[h]
%\hspace{2mm}
%\begin{minipage}{0.48\linewidth}
%\hspace{-2mm}
\includegraphics[width=.88\linewidth]{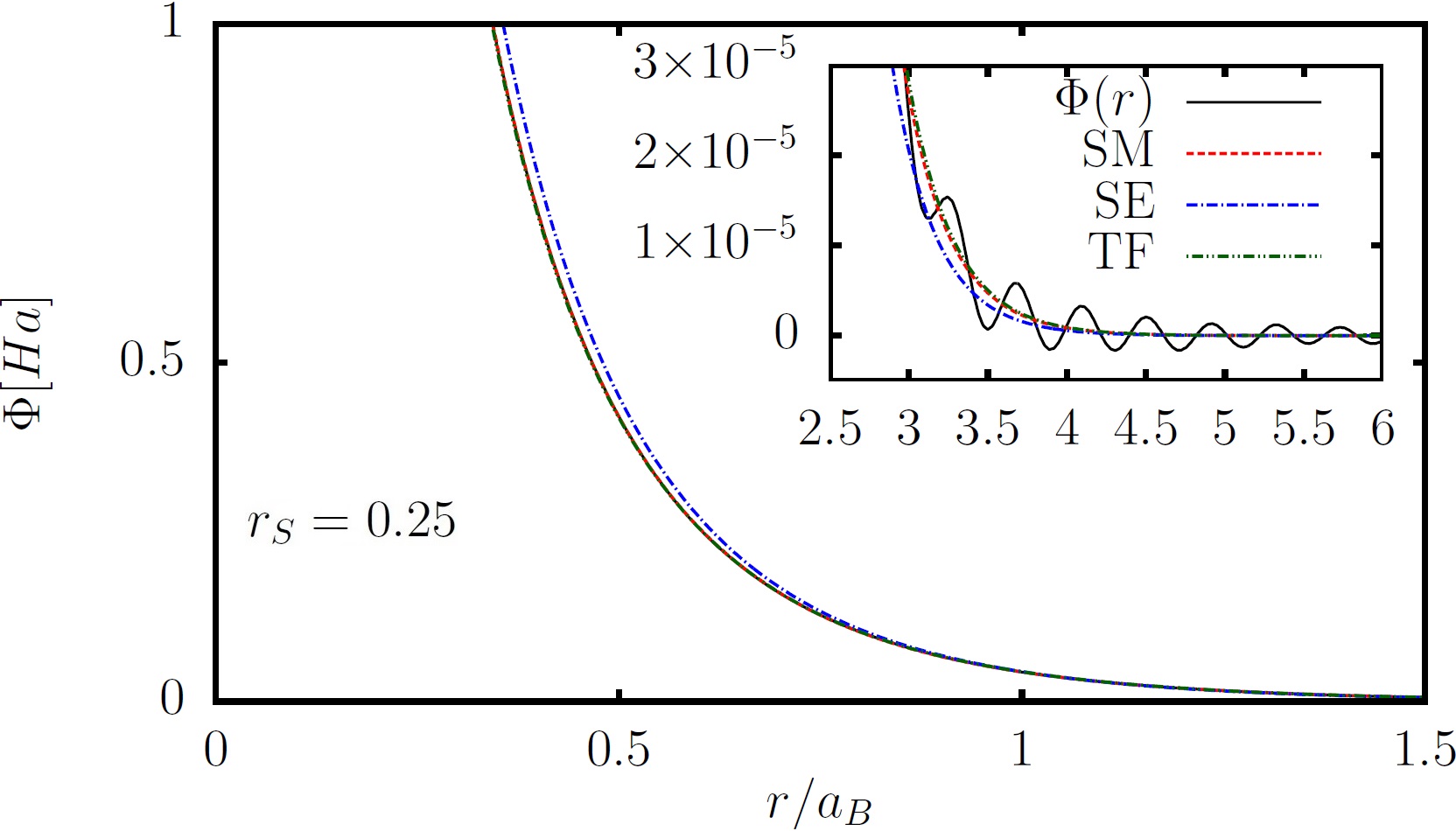}\\[2ex]
\includegraphics[width=.88\linewidth]{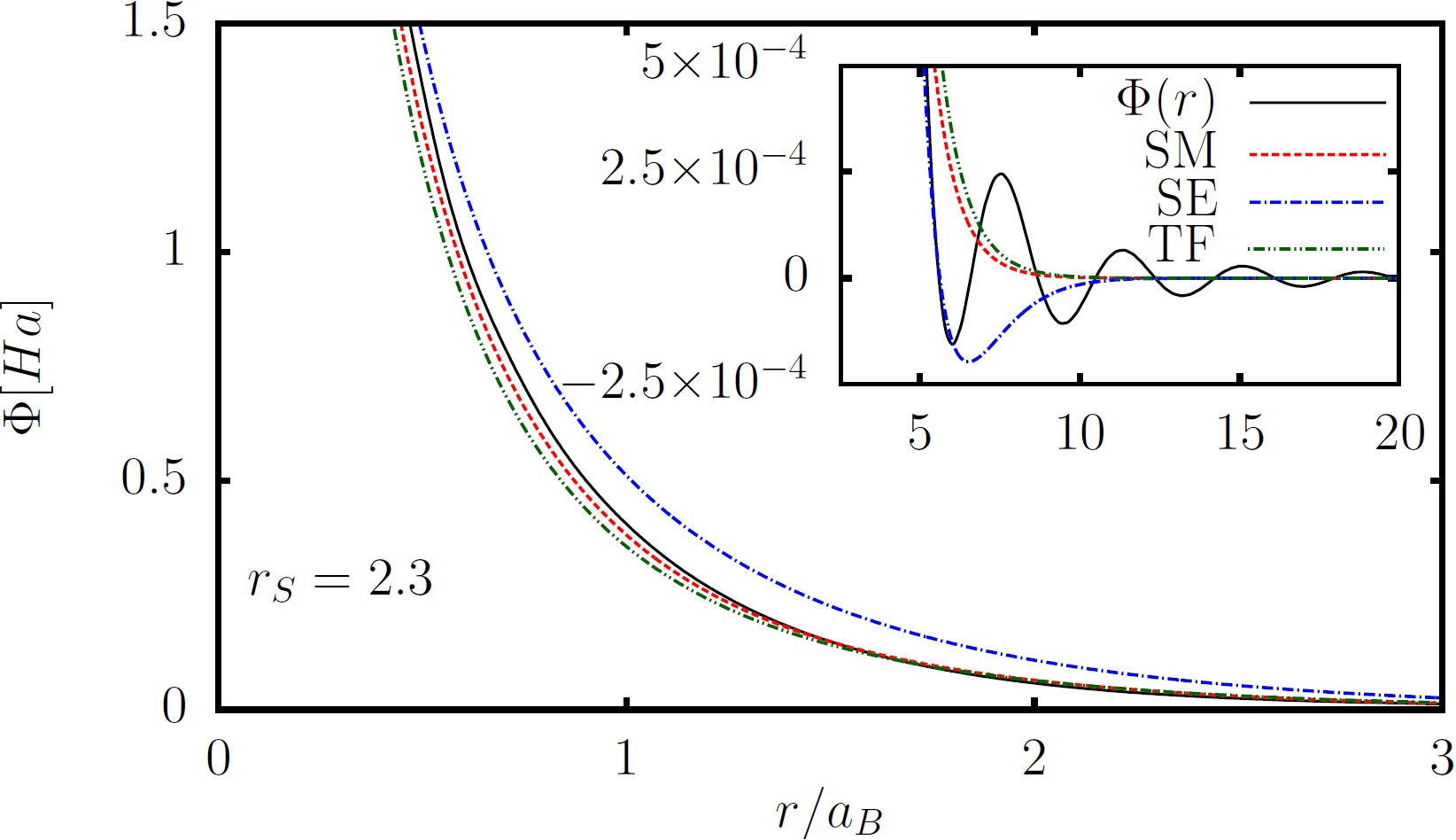}\\[2ex]
\includegraphics[width=.88\linewidth]{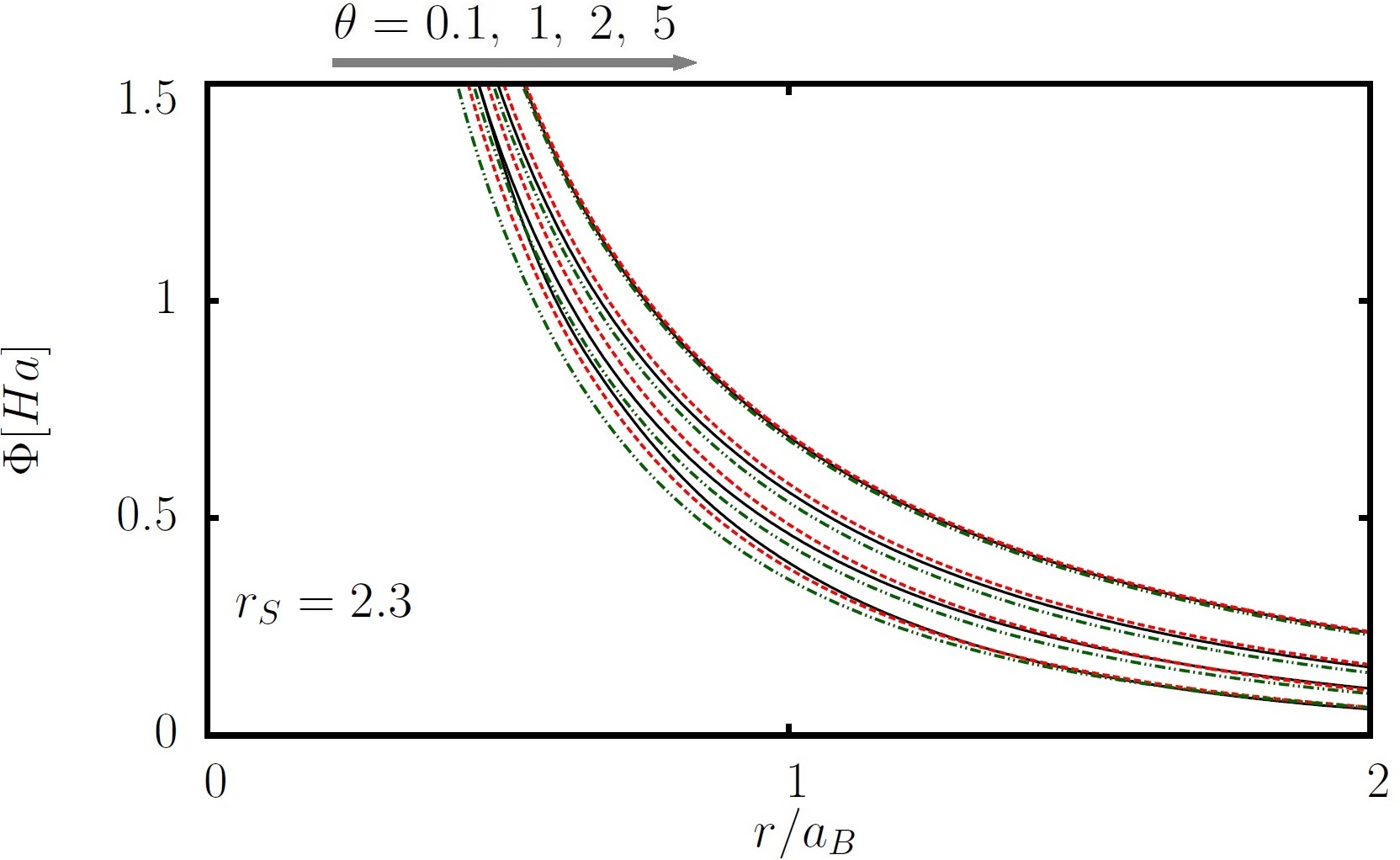}
\caption{(Color online) Top two figures: same as middle and bottom parts of Fig.~\ref{fig:all_vs_rpa0} but with the Yukawa potential, Eq.~(\ref{Yukawa}), ``TF'', added for comparison of the ground state behavior. Bottom: temperature dependence of the potentials. Same as bottom part of Fig.~\ref{fig:ms_rpa_t} with the Yukawa potential (\ref{Yukawa}) added.}
\label{fig:all_vs_rpa0_tf}
\end{figure}

To compare the first and second approximations to the static inverse dielectric function, Eqs.~(\ref{eq:eps_0}) and (\ref{eq:eps_2}), we plot again the associated static potentials computed according to Eq.~(\ref{POT_stat}). In the top two parts of Fig.~\ref{fig:all_vs_rpa0_tf} we show again the zero temperature results of SE and SM and, in addition the Thomas-Fermi screended Yukawa result, Eq.~(\ref{Yukawa}). Thus, we are able to assess the quality of the different orders of the wavenumber expansion of $\epsilon^{-1}$, in comparison to the full RPA result. For $r_s=2.3$ we see that the Yukawa potential (zeroth order of the expansion) provides a good approximation, but the next correction, proportional to $k^2$, that is contained in Eq.~(\ref{eq:eps_2}) and, thus, in the SM potential constitutes an improvement. In contrast, the SE potential which contains the $k^2$ correction with a nine times larger prefactor exhibits a poor performance and is even significantly less accurate than the zeroth order correction alone. The same trend is observed at higher density (top figure) where the zeroth and first order are both indistinguishable from the full result whereas the SE potential is substantially less accurate. 

The bottom part of Fig.~\ref{fig:all_vs_rpa0_tf} shows the performance of the zeroth and second order potentials [TF and SM, Eqs.~(\ref{eq:eps_0}) and (\ref{eq:eps_2}), respectively] at finite temperature. While at low temperature, $\Theta=0.1$, the second order provides a significant impovement, for higher temperature, $\Theta \gtrsim 1$, the expansion exhibits sign alternating convergence and 
both approximations are of comparable quality with a slightly better performance of the SM potential. In other words, for $\Theta \gtrsim 1$, the standard static Yukawa potential provides a fairly accurate description of the static limit of the RPA.

{\bf Higher order results. Correlations:} It poses no principal problem to extend the above results for the inverse dielectric function to higher powers of $k$. With the equations (\ref{eq:a4}--\ref{eq:a8}) the next three higher orders are available, for arbitrary temperature. This should improve the behavior of the static potential also at shorter distances. However, for the short-distance behavior also other effects will be important, most noticeably exchanges and correlations. One way to include correlations is to use static local field corrections, as proposed in Ref.~\cite{stanton_pre15}. A more systematic and powerful approach is to use many-body theory which provides dynamic correlation corrections, e.g. Refs. \cite{kwong_prl_00} and \cite{bonitz-book} and the cited references.

\subsection{Conclusions}
The conclusion from the above comparison of approximate static ion potentials to the RPA result is that, in fact, there is no freedom in the choice of the prefactor of the Bohm term for quantum hydrodynamics for fermions. The QHD equations follow from the Thomas-Fermi theory by performing a systematic gradient expansion of the kinetic energy or, equivalently, a long-wavelength expansion of the inverse polarization function ${\tilde K}$, Eq.~(\ref{eq:pinv_exp}). This expansion directly leads to the coefficient $1/9$, at $T=0$, in front of the Bohm term compared to the cases of bosons or of a single-particle, whereas for higher temperatures the coefficient increases monotonically to $0.3$ (in 1D and 2D different coefficients apply \cite{michta_cpp15}).
This coefficient is a consequence of the Pauli principle and has important implications for a hydrodynamic modeling of dense quantum plasmas. For example, we have shown that a neglect of this coefficient gives rise to a statically screened ion potential that is far less accurate than the standard Thomas-Fermi (Yukawa) potential, rather than an improvement.  It also indicates that earlier derivations of QHD equations 
 are invalid for quantum plasmas and earlier QHD results that were using the prefactor $1$ do not describe fermions, in general, nor dense plasmas, in particular.

\subsection*{Acknowledgments}

This work has been supported by the Deutsche Forschungsgemeinschaft via grant SFB-TR 24, project A9 and  the Ministry of Education and Science of
Kazakhstan.

\section*{Appendix: Finite temperature gradient corrections of the kinetic energy and inverse polarization}\label{s:A}
Here we present a systematic derivation of the gradient correction to the Thomas Fermi energy using Mermin's approach for finite temperatures. This appears to be the most systematic and general approach and allows one to derive higher order corrections as well.
\subsection*{Gradient expansion of the kinetic energy}
Following Hohenberg and Kohn \cite{Kohn}, and Mermin \cite{Mermin2}, the electron energy is a functional of the density:
\begin{equation}\label{En}
E[n]=\int \! v(\vec{r})n(\vec{r})\, \mathrm{d}\vec{r}+\frac{1}{2}\int \! \frac{n(\vec{r})n(\vec{r}^\prime)}{\mid \vec{r}-\vec{r}^\prime\mid} \, \mathrm{d}\vec{r}\mathrm{d}\vec{r}^\prime+
T[n],
\end{equation}
where the first term is the energy due to external field and the second is the interaction energy of the electrons in mean field
approximation. $T[n]$ is the kinetic energy (we neglect additional exchange-correlation terms) which is well known in the local approximation (Thomas-Fermi theory),
\begin{equation}
 T_0[n] 
% = T[n_0] 
 = \int d{\bf r} \, t[n({\bf r})]\, n({\bf r}),
\end{equation}
where the kinetic energy density is related to the (local) Fermi energy, $t[n]=\frac{3}{5}E_F[n]$.
To obtain corrections to this local approximation, we expand the kinetic energy in terms of the density perturbation, $\tilde{n}=n(\vec{r})-n_0$, 
\begin{multline}\label{Gn}
T[n]=T_0[n_0]+\int \! K(\vec{r}-\vec{r}^\prime)\tilde{n}(\vec{r})\tilde{n}(\vec{r^{\prime}})\, \mathrm{d}\vec{r}\mathrm{d}\vec{r}^\prime+\\
+\int \! L(\vec{r},\vec{r}^\prime,\vec{r}^{\prime \prime})\tilde{n}(\vec{r})\tilde{n}(\vec{r^{\prime}})\tilde{n}(\vec{r^{\prime \prime}})\, \mathrm{d}\vec{r}\mathrm{d}\vec{r}^\prime\mathrm{d}\vec{r}^{\prime \prime}+....
\end{multline}
%here  $n(\vec{r})=n_0+\tilde{n}(\vec{r})$ was introduced:
%\begin{equation}\label{dn}
%where $\int \! \tilde{n}(\vec{r})\, \mathrm{d}\vec{r}=0.$
%\end{equation}
%
%Formula~(\ref{Gn}) allows us to obtain corrections to the Thomas-Fermi kinetic energy, $T[n_0]$. 
We now transform the second term to an expansion in powers of the density gradient (gradient corrections),
%Further we consider the gradient correction [the second term in (\ref{Gn})], 
by Fourier expanding the kernel ($\Omega$ is the volume), 
\begin{equation}\label{K}
K(\vec{r})=\frac{1}{\Omega} \sum_{\vec{k}} {\tilde K}(\vec{k})\,e^{-i \vec{k}\cdot \vec{r}}.
\end{equation}
From Eqs.~(\ref{Gn}) and (\ref{K}) we have, using the convolution theorem,
\begin{equation}\label{Gn2}
T[n]=T_0[n_0]+\frac{1}{\Omega} \sum_{\vec{k}} {\tilde K}(\vec{k})\tilde{n}(\vec{k})^2 + \dots.
\end{equation}
We are now looking for the long-wavelength limit and expand
\begin{equation}\label{K3}
{\tilde K}(\vec{k})={\tilde a}_0
%+{\tilde a}_1 \vec{k}
+{\tilde a}_2 \vec{k}^2 + {\tilde a}_4 \vec{k}^4+...,
\end{equation}
Since $K(\vec{r})$ is real, Eq.~(\ref{K}) leads to ${\tilde K}(-\vec{k})={\tilde K}(\vec{k})$, therefore, the expansion (\ref{K3}) contains 
only even powers of ${\bf k}$.
From this, we obtain for $K(\vec{r}-\vec{r}^{\prime})$, using (\ref{K}),
\begin{equation}\label{K4}
K(\vec{r}-\vec{r}^{\prime})=({\tilde a}_0
%+i{\tilde a}_1\nabla 
- {\tilde a}_2 \nabla \nabla^{\prime}+...)\delta (\vec{r}-\vec{r}^{\prime}).
\end{equation}
Substituting (\ref{K4}) into (\ref{Gn}) one obtains
\begin{align}\label{Gn4}
T[n] &= T_0[n_0] + \sum_{l=1}  T_{2l}[n],
\\
T_{2l}[n] &=  \int \mathrm{d}\vec{r}\,a_{2l}\left[n\right] \mid \nabla^l n(\vec{r})\mid^{2}, 
\label{term_2l}
%T[n_0]-\int \! a_2 \mid \nabla n(\vec{r})\mid^{2} \, \mathrm{d}\vec{r}+...,
\end{align}
where we took into account that $\int \! \tilde{n}(\vec{r})\, \mathrm{d}\vec{r}=0$ 
%and the fact that odd powers of the density gradient cannot contribute to the kinetic energy, $a_1 = a_3 = \dots 0$. 
and identifiy $a_2 = - {\tilde a}_2$.

Thus we have rewritten the kinetic energy in terms of a local term plus gradient corrections. It was shown in Ref.~\cite{michta_cpp15} that this energy functional leads directly, via functional differentiation with respect to the density profile, to quantum hydrodynamic equations with the local term yielding the Fermi energy and the first gradient term giving rise 
to a Bohm-type potential,
\begin{equation}
  \frac{\delta T_{2}[n;\gamma]}{\delta n} = \gamma\, \frac{\hbar^2}{8m}\left( \left| \frac{\nabla n}{n} \right|^2 - 2 \frac{\nabla^2 n}{n} \right) 
 = \gamma \,V^{\rm MH}_B[n],
\label{eq:grad=vb}
\end{equation}
where $V^{\rm MH}_B$ is the potential postulated by Manfredi and Haas \cite{manfredi-haas01}. What remains is to establish the
relation between the coefficient $\gamma$ and the coeffient $a_2$ of the expansion (\ref{term_2l}). In the following we prove that, for zero temperature, to lowest order in the correlation energy, $a_2 = \frac{\hbar^2}{72 m}$ which means that $\gamma = 1/9$.

%------------
\subsection*{Finite temperature first order gradient term in RPA}
To evaluate the coefficient $a_2 = - {\tilde a}_2$ we recall that the energy associated with a charged particle density fluctuation can be expressed, in linear response, in terms of an effective potential (external plus induced potential) 
%leads to an additional energy due to ion-electron interaction
%
\begin{equation}\label{Gn3}
T[n]=T_0[n_0]+\frac{1}{2\Omega} \sum_{\vec{k}}\tilde{n}(\vec{k}) \tilde{U}_{\rm eff}(\vec{k}).
\end{equation}
Recalling the definition of the longitudinal polarization function, $\Pi(\vec{k})=\tilde{n}(\vec{k})/\tilde{U}_{\rm eff}(\vec{k})$,
Eqs.~(\ref{Gn2}) and (\ref{Gn3}) allow us to identify
\begin{equation}\label{K2}
{\tilde K}(\vec{k})=\frac{1}{2\Pi(\vec{k})},
\end{equation}
and to make use of known results for $\Pi$ below.
The lowest order many-body approximation for $\Pi$ is the random phase approximation (RPA) which reads, for an arbitrary temperature \cite{Arista} 
\begin{equation}\label{Pi}
\Pi_{\rm RPA}(k,\omega)=-\frac{k^2\chi_{0}^{2}}{16 \pi e^2z^3}\left[g(u+z)-g(u-z)\right],
\end{equation}
where  $u=\omega/(kv_F)$, $z=k/(2k_F)$, $\chi_{0}^{2}=3/16 \left( \hbar\omega_p/E_F\right)^2$, $k_F=(3\pi^2n)^{1/3}$, $\omega_{p}^{2}=4\pi ne^2/m_e$, and
\begin{equation}\label{g}
g(x)=-g(-x)=\int \! \frac{y\,\mathrm{d}y}{\exp(y^2/\Theta-\eta)+1}\ln\left|\frac{x+y}{x-y}\right|,
\end{equation}
where $\Theta=k_BT/E_{F}$, and $\eta=\mu/k_BT$ is the chemical potential.
To obtain the long-wavelength limit of ${\tilde K}$, Eq.~(\ref{K3}), we now expand the inverse of the static (i.e. $\omega=u=0$) polarization function in the limit $z\ll 1$:

\begin{align}\label{K5}
{\tilde K}_{\rm RPA}(k) &= -\frac{\pi e^2}{2k_F^2\chi_0^2 H_1(\eta)}+\frac{\pi e^2H_2(\eta)}{24k_F^4\chi_0^2 H_1^2(\eta)}\cdot k^2,
\\[1ex]
H_1(\eta) &= \frac{\sqrt{\Theta}}{2} \, I_{-1/2}(\eta), \quad H_2(\eta)=\frac{1}{2\sqrt{\Theta}} \,I_{-3/2}(\eta).
\end{align}
%here $H_1(\eta)=(4D)^{-1/2} I_{-1/2}(\eta)$, and $H_2(\eta)=D^{1/2}I_{-3/2}(\eta)/2$.
The first term in (\ref{K5}) yields the coefficient ${\tilde a}_0$ in Eq.~(\ref{K3}) which does not contribute to the kinetic energy since the integral of ${\tilde n}$ vanishes whereas the second term in (\ref{K5}) yields the coefficient $a_2$:
\begin{align}\label{a2}
a_0\left(\left[n\right],\Theta\right) &=-\frac{\pi e^2}{2k_F^2\chi_0^2 H_1(\eta)},
\\
a_2\left(\left[n\right],\Theta\right) &=-\frac{\hbar^2 H_2(\eta)}{72m_e n H_1^2(\eta)}.
\end{align}
In the ground state, $\Theta\rightarrow 0$, we have the limits $H_1(\eta)\simeq 1$ and $H_2\simeq -1$,
so $H_2/H_1^2 \simeq -1$, and  (\ref{a2}) gives $a_2(0) \to \hbar^2/(72m)$ meaning that the coefficient in front of the Bohm potential in Eq.~(\ref{eq:grad=vb}) becomes $\gamma = 1/9$. The temperature dependence of this coefficient was displayed in Fig.~\ref{fig:alpha_3}.

Finally, substituting $a_2$ from (\ref{a2}) to the formula (\ref{Gn4}), for the kinetic energy at finite temperature we have \cite{Perrot, kirzhnits75}:
\begin{equation}\label{Gn5}
T[n]=T_0[n_0]+ \frac{1}{9} \frac{\hbar^2}{8m_e} 4\beta^{3/2} \int \! E_{F}^{3/2}(n)\frac{I_{-1/2}^{\prime}(\eta)}{I_{-1/2}^2(\eta)} \frac{\mid \nabla n\mid^{2}}{n} \, \mathrm{d}\vec{r},
\end{equation}
were we eliminated the Fermi integral of order $-3/2$ using the formula $I_{-3/2}=-2I_{-1/2}^{\prime}$.
  
%------------
\subsection*{Higher order gradient terms in RPA for finite $T$}
Now consider the higher order ($l > 1$) terms of the gradient expansion (\ref{Gn4}). In the long-wavelength limit, $z<1$ the kernel ${\tilde K}$ can be written as:
\begin{equation}\label{K6}
K_{\rm RPA}(k)=-\frac{\pi^2\hbar^2 \Theta^{-1/2} }{m_e k_F I_{-1/2}(1+\sum_i b_i z^{2i})},
\end{equation}
where
\begin{equation}
\label{eq:bi}
 b_i\left(\left[n\right],\Theta\right)=\frac{\Theta^{-i}}{2i+1} \frac{I_{-i-1/2}(\eta)}{I_{-1/2}(\eta)}.
%\nonumber
\end{equation}
Expanding (\ref{K6}) it is straightforward to obtain the higher (even) order terms of the gradient correction:
\begin{equation}
%\label{GGG1}
T_4[n]=\frac{\hbar^2 \pi^2}{16 m_e }  \int \! \frac{b_1^2-b_2}{\Theta\left[n\right]^{1/2}k_F^5I_{-1/2}(\eta)}\mid \nabla^2 n\mid^{2} \, \mathrm{d}\vec{r},
\nonumber
\end{equation}
\begin{equation}
%\label{GGG2}
T_6[n]=\frac{\hbar^2 \pi^2}{64 m_e }  \int \! \frac{b_1^3-2b_1b_2+b_3}{\Theta\left[n\right]^{1/2}k_F^7I_{-1/2}(\eta)}\mid \nabla^3 n\mid^{2} \, \mathrm{d}\vec{r},
\nonumber
\end{equation}
\begin{multline}
%\label{GGG4}
T_8[n]=\frac{\hbar^2 \pi^2}{256 m_e}\times \\
 \int \! \frac{b_1^4-3b_1^2b_2+b_2^2+2b_1b_3-b_4}{\Theta\left[n\right]^{1/2}k_F^9I_{-1/2}(\eta)}\mid \nabla^4 n\mid^{2} \, \mathrm{d}\vec{r}.
\nonumber
 \end{multline}
In the zero-temperature limit, $T_4$ and $T_6$ were obtained by Hodges \cite{Hodges} and Murphy \cite{Murphy}, respectively. 
For finite temperature these terms and,  additionally, $T_8$ are given here for the first time. 

Here we do not discuss the convergence of this gradient series. Obviously, before using higher order terms one should verify that all terms of the same order arising from three-particle correlations [third term in (\ref{Gn})] are included as well. Further, it is well possible that there exist comparable terms arising from correlation effects beyond the RPA, e.g. \cite{kwong_prl_00}.

Finally we note that these results apply only to three-dimensional plasmas. Results for 1D and 2D systems were discussed in Ref.~\cite{michta_cpp15}. Extensions of these results to finite temperatures are given in Ref.~\cite{zhandos_cpp_new}.


\begin{thebibliography}{10}
\bibitem{kpmnote} In the original formulas $k'_\pm$ are denoted by $k_r$ and $k_i$, respectively.
\end{thebibliography}
\end{document}